\documentclass[11pt]{article}
\usepackage{amsmath}
\usepackage{amsfonts}
\usepackage{amssymb}
\usepackage{bbm}
\usepackage{latexsym}
\usepackage[dvips]{epsfig}
\usepackage{graphicx}
\usepackage{psfrag}
\usepackage{pstricks}
\usepackage{colordvi}
\usepackage{subfig}
\usepackage{wrapfig}
\usepackage{amsmath,amsfonts,amssymb}
\usepackage{mathrsfs}
\usepackage{url}
\usepackage{array}

\makeatletter\@addtoreset{equation}{section}\makeatother

\newcommand{\bb}{\mbox{${\bf b}$}}

\newcommand{\bI}{\mbox{\boldmath{$I$}}}

\newcommand{\bn}{\mbox{${\bf n}$}}

\newcommand{\bp}{\mbox{${\bf p}$}}
\newcommand{\bP}{\mbox{${\bf P}$}}

\newcommand{\bQ}{\mbox{${\bf Q}$}}
\newcommand{\bq}{\mbox{${\bf q}$}}

\newcommand{{\bs}}{\mbox{${\bf s}$}}
\newcommand{\bt}{\mbox{\boldmath{$t$}}}

\newcommand{{\bS}}{\mbox{${\bf S}$}}
\newcommand{\bu}{\mbox{${\bf u}$}}

\newcommand{\bv}{\mbox{${\bf v}$}}

\newcommand{\bx}{\mbox{${\bf x}$}}

\newcommand{\by}{\mbox{${\bf y}$}}

\newcommand{\bepsilon}{\mbox{\boldmath{$\varepsilon$}}}

\newcommand{\bsigma}{\mbox{\boldmath{$\sigma$}}}

\newcommand{\bpi}{\mbox{\boldmath{$\pi$}}}
\newcommand{\bPi}{\mbox{\boldmath{$\Pi$}}}

\newcommand{\bzero}{\mbox{$\bf 0$}}

\newcommand{\BBC}{\mbox{$\mathbb{C}$}}

\newcommand{\BBR}{\mbox{$\mathbb{R}$}}

\newfont{\twelvemsb}{msbm10 at 11.6pt}

\newcommand{\half}{\mbox{$\frac{1}{2}$}}

\renewcommand{\div}{\mathop{\rm div}\nolimits}

\newcommand{\tr}{\mathop{\rm tr}}

\newcommand{\beqn}{\begin {equation}}

\newcommand{\dfracp}[2]{\dfrac{\partial #1}{\partial#2}}

\newcommand{\fig}[1]{Fig.~\ref{#1}}

\renewcommand{\b}[1]{\boldsymbol{#1}} 

\renewcommand{\o}[1]{\overline{#1}}

\renewcommand{\d}{\mathsf{d}}

\renewcommand{\div}{\mbox{div}}

\renewcommand{\tr}{\mbox{tr}}

\definecolor{gray}{rgb}{0.75, 0.75, 0.75}
\definecolor{yellow}{rgb}{1, 0.7, 0.2}
\definecolor{green}{rgb}{0.3, 0.9, 0.3}
\definecolor{brown}{rgb}{0.6, 0.3, 0.2}
\definecolor{magenta}{rgb}{0.9, 0.1, 0.9}
\definecolor{light}{rgb}{1, 0.7, 0.7}
\usepackage{siunitx}
\usepackage{multirow}

\begin{document}
\begin{center}
{\Large {\sc Some Properties of the Dissipative Model}}\\
{\Large {\sc of Strain-Gradient Plasticity}}
\vspace{3ex}\\
C Carstensen$^1$, F Ebobisse$^{2}$, AT McBride$^{3}$, BD Reddy$^{\star,2}$, P Steinmann$^{4}$
\let\thefootnote\relax\footnote{$^\star$ Corresponding author} 
\let\thefootnote\relax\footnote{$^1$ Institut f\"ur Mathematik, Humboldt-Universit\"at zu Berlin, Unter den Linden 6, D-10099 Berlin, Germany. E-mail cc@math.hu-berlin.de} 
\let\thefootnote\relax\footnote{$^2$ Department of Mathematics and Applied Mathematics, University of Cape Town, 7701 Rondebosch, South Africa. Email \{francois.ebobissebille,daya.reddy\}@uct.ac.za } 
\let\thefootnote\relax\footnote{$^3$ School of Engineering, The University of Glasgow, Glasgow G12 8QQ, United Kingdom. Email
andrew.mcbride@glasgow.ac.uk} 
\let\thefootnote\relax\footnote{$^4$ Chair of Applied Mechanics, University of Erlangen-Nuremberg, Egerlandstr. 5, 91058 Erlangen, Germany. Email  paul.steinmann@ltm.uni-erlangen.de} 
\end{center}
{\small \section*{Abstract} 
A theoretical and computational investigation is carried out of a dissipative model of rate-independent strain-gradient plasticity and its regularization. It is shown that the flow relation, when expressed in terms of the Cauchy stress, is necessarily global. The most convenient approach to formulating the flow relation is through the use of a dissipation function. It is shown, however, that the task of obtaining the dual version, in the form of a normality relation, is a complex one. A numerical investigation casts further light on the response using the dissipative theory in situations of non-proportional loading. The elastic gap, a feature reported in recent investigations, is observed in situations in which passivation has been imposed. It is shown computationally that the gap may be regarded as an efficient path between a load-deformation response corresponding to micro-free boundary conditions, and that corresponding to micro-hard boundary conditions, in which plastic strains are set equal to zero.
}

\section{Introduction}
There has been steady progress in the development of strain-gradient theories of plasticity for over two decades, since the early contribution by Aifantis \cite{Aifantis}. The motivation for such theories lies in their ability to capture length-scale dependent effects, which conventional theories are unable to do. Some key works include those by Gao, Huang, Nix and Hutchinson \cite{GHN,GHNH}, who argue for the inclusion of gradients of plastic strain as a way of accounting for geometrically necessary dislocations, and Fleck and Hutchinson, Gudmundson, and Gurtin and Anand \cite{FH,Gudmundson,GA}, who develop such theoretical models. 

This work concerns the small-strain, rate-independent theory of strain-gradient plasticity. The model is based on that first proposed for rate-dependent materials by Gurtin and Anand \cite{GA}, and subsequently developed for the rate-independent case in \cite{R,REM}. These authors also carried out an analysis of well-posedness of the problem. The works by Fleck and Willis \cite{FW1,FW2} present and analyse closely related rate-independent and -dependent theories. 

In the models referred to above, gradient effects are accounted for either through their inclusion in the free energy, or in an extension of the flow law. These are referred to respectively as energetic and dissipative models, and are both present in many treatments of gradient plasticity. They differ substantially though in their implications for the theory. Fleck, Hutchinson and Willis \cite{FHW1}, for example, point out that it is particularly in cases of non-proportional loading that the energetic and dissipative models lead to quite distinct behaviour. These authors refer to these respectively as incremental and non-incremental theories: their nomenclature stems from the observation that, for energetic (or incremental) theories it is possible to express increments in the microscopic stresses that form part of the description of the model in terms  of increments in plastic strain and strain gradients. On the other hand, at least when expressed in local form in a manner that mimics the classical associative flow law, the dissipative model leads to the expression of microscopic stresses -- not their increments -- in terms of plastic strain and strain gradient  increments. These differences in the models are explored and highlighted in \cite{FHW1} in analyses of two problems that involve non-proportional loading. The main distinguishing feature in the two examples is, in the case of the dissipative theory, an elastic gap: that is, elastic behaviour associated with non-proportional loading following loading into the plastic range. This phenomenon has been further investigated in \cite{FHW2}.

The yield criterion and associative flow law for the strain-gradient problem gives the plastic strain-rate (or increment) and its gradient in terms of a normality condition that involves the yield function as a function of the microscopic stresses. Unlike the Cauchy stress these are not known a priori in terms of current displacement and plastic strain and therefore cannot be used to determine whether yield has occurred locally,  as has been discussed in \cite{FW1,FW2}.  It has been shown in \cite{R} however that the microstresses can be eliminated in favour of the Cauchy stress in the flow relation by resorting to a weak or global form of the flow law. This global form is most conveniently written in terms of the dissipation function, from which the flow relation as a normality law can in principle be obtained via a dualization procedure.

The objective of this work is to explore various aspects of the dissipative strain-gradient theory, with a view to shedding further light on features of the theory that include those explored in \cite{FHW1,FHW2}. We summarize the relevant governing relations in Section 2 and derive the flow law in global form, in terms of the dissipation function and involving the Cauchy stress. A mixed formulation, obtained by introducing an auxiliary variable for the plastic strain gradient, is presented in Section 2.2.
  
 Section 3 explores the implications of a regularized theory. The dissipation function is not smooth at the origin, and is approximated in Section 3 by one that is smooth. One consequence is that inequalities corresponding to the flow relations are replaced by local or global equations. 

In Section 4 time-discretization allows the global flow relation to be formulated as one involving plastic strain increments, and for the problem to be formulated as a minimization problem. Such a formulation is not possible for the original problem.  

In Section 5 we approach the issue of finding the yield function by replacing the original global problem with its fully (spatially and temporally) discrete approximation. Remarkably, even for the discrete problem it is possible only to find an upper bound for the yield function, as shown in Section 5.2. 

Section 6 is devoted to a numerical investigation of the problem, with the focus on non-proportional loading. The problem discussed in \cite{FHW1} of a strip in tension is revisited, with an alternative explanation for the occurrence of the elastic gap. We then discuss two further problems, viz. biaxial deformation of a thin plate, and extension of a circular cylindrical rod. Non-proportional loading is effected through  a change in loading direction and through the application of passivation, that is, imposition of zero plastic strain increment on part of the boundary. 

The elastic gap reported in \cite{FHW1} is observed in situations in which passivation has been imposed. An interpretation, from a mathematical perspective, of the gap is given by appealing to the expression for the yield function as a maximum, taken over all admissible plastic strain increments, of a function involving the dissipation. Numerically, the elastic gap appears to constitute an efficient transition from stress-strain behaviour corresponding to a micro-free or Neumann boundary condition, to that which is obtained assuming micro-hard or Dirichlet boundary conditions. 

\section{Governing equations and inequalities}\label{sec:gov}
The model of strain-gradient plasticity that forms the basis of this study is that proposed by Gurtin and Anand \cite{GA}, with the specialization to rate-independent plasticity by Reddy \cite{R}. Small strains are assumed. The displacement is denoted by $\bu$, the total strain by $\bepsilon$, and the stress by $\bsigma$. Small strains are assumed. The strain is decomposed into elastic and plastic components $\bepsilon^e$ and $\bepsilon^p$ according to 
\begin{equation}
\bepsilon = \bepsilon^e + \bepsilon^p\,.
\label{straindecomp}
\end{equation}
The strain-gradient theory makes provision for a 2nd-order microscopic stress tensor $\bpi$ and a 3rd-order microscopic stress $\bPi$. The quantity $\bpi$ is symmetric and deviatoric, while $\bPi$ is symmetric and deviatoric in its first two indices, in the sense that $\Pi_{ijk} = \Pi_{jik},\ \Pi_{ppk} = 0$. Here and elsewhere the summation convention on repeated indices is invoked, with partial derivatives denoted by a subscript following a comma.

We define the generalized stress ${\sf S}$ and plastic strain ${\sf \Gamma}$ to be the ordered pairs
\begin{equation}
{\sf S} = (\bpi,\ell^{-1}\bPi),\qquad
{\sf \Gamma} = (\bepsilon^p, \ell \nabla \bepsilon^p)\,.
\label{SandGamma}
\end{equation}
Here $\ell$ is a length parameter, and the inner product of the two generalized quantities is denoted by
\[
{\sf S}\diamond {\sf \Gamma}
  := \bpi:\bepsilon^p + \bPi\circ\nabla\bepsilon^p = \pi_{ij}\varepsilon^p_{ij} 
  + \Pi_{ijk}\varepsilon^p_{ij,k}\,.
\]
Assuming quasistatic behaviour, the equation of macroscopic equilibrium is given by 
\begin{equation}
- \mbox{div}\,\bsigma = \bb \,,
\label{equm}
\end{equation}
where $\bb$ is the body force. In addition, the stress and microscopic stresses are related to each other through the microforce balance equation
\begin{equation}
\mbox{dev}\,\bsigma = \bpi - \mbox{div}\,\bPi\quad\mbox{or, in index form,}\quad (\mbox{dev}\,\bsigma)_{ij} = \pi_{ij} - \Pi_{ijk,k}\,.
\label{mfb}
\end{equation}
Equations \eqref{equm} and \eqref{mfb} are required to be satisfied on the domain $\Omega$.
The macroscopic boundary conditions on the problem are
\begin{equation}
\bu = \bar{\bu}\ \mbox{on}\ \partial \Omega_u\,,\qquad \bsigma\bn = \bar{\bt}\ \mbox{on}\ \partial \Omega_t\,,
\label{macrobcs}
\end{equation}
where $\partial \Omega_u$ and $\partial \Omega_t$ are complementary parts of the boundary $\partial \Omega$ with unit outward normal $\bn$, and $\bar{\bu}$ and $\bar{\bt}$ are respectively a prescribed displacement and surface traction. In addition we assume homogeneous micro-hard and micro-free boundary conditions on complementary parts $\partial \Omega_H$ and $\partial \Omega_F$ of the boundary; that is,  
\begin{equation}
\bepsilon^p = \bzero\ \mbox{on}\ \partial \Omega_H\,,\qquad \bPi\bn = \bzero\ \mbox{on}\ \partial \Omega_F\,.
\label{microbcs}
\end{equation}

Of particular interest is the weak form of the microforce balance equation \eqref{mfb}. We denote by $W$ the set of plastic strains, defined by 
\[
W = \{ \bq\ |\ q_{ij} = q_{ji},\ q_{ii} = 0,\ q_{ij} \in L^2(\Omega),\ q_{ij,k} \in L^2(\Omega),\ q_{ij} = 0\ \mbox{on}\ \partial \Omega_H\}\,.
\] 
Taking the inner product of \eqref{mfb} with arbitrary $\bq \in W$, integrating by parts, and imposing the microscopic boundary conditions \eqref{microbcs}, we obtain the weak formulation 
\begin{align}
\int_\Omega \mbox{dev}\,\bsigma:\bq\ dx & = \int_\Omega [\bpi:\bq + \bPi\circ\nabla\bq ] \ dx \nonumber \\
& = \int_\Omega {\sf S} \diamond {\sf Q}\ dx\,,
\label{mfbweak}
\end{align}
where ${\sf Q} = (\bq,\ell\nabla \bq)$.

Given the free energy $\psi$ the free-energy imbalance takes the form
\begin{equation}
\dot{\psi} - \bsigma:\dot{\bepsilon}^e - \bpi:\dot{\bepsilon}^p -\bPi\circ\nabla\dot{\bepsilon}^p \leq 0\,.
\label{dissineq}
\end{equation}
Since we are concerned in this work with the consequences of a dissipative gradient plasticity formulation we restrict attention to free energy functions of the form\footnote{More generally, one considers a free energy that depends in addition on the plastic strain, the plastic strain gradient and, possibly, hardening internal variables. Details may be found, for example, in \cite{R}.}
\begin{equation}
\psi = \psi^e(\bepsilon^e) = \half \bepsilon^e:\BBC\bepsilon^e \,,
\label{psie}
\end{equation}
in which the elasticity tensor $\BBC$ is given, for isotropic materials, by
\begin{equation}
\BBC\bepsilon = \lambda (\mbox{tr}\,\bepsilon)\bI + 2\mu\,\bepsilon\,.
\label{C}
\end{equation}
Here $\lambda$ and $\mu$ are the Lam\'{e} parameters. 
We note also for future reference that the deviatoric part of this relation is given by
\begin{equation}
\mbox{dev}\,\BBC\bepsilon = 2\mu\,\mbox{dev}\,\bepsilon\,.
\label{Cdev}
\end{equation}
Substitution of \eqref{psie} in \eqref{dissineq} and the usual Coleman-Noll procedure lead to the elastic relation
\begin{equation}
\bsigma= \frac{\partial \psi^e}{\partial \bepsilon^e} = \BBC\bepsilon^e
\label{elastreln}
\end{equation}
and the reduced dissipation inequality
\begin{equation}
\bpi:\dot{\bepsilon}^p + \bPi\circ\nabla\dot{\bepsilon}^p 
\geq 0
\quad \mbox{or}\quad {\sf S} \diamond \dot{\sf \Gamma} 
\geq 0\,.
\label{reddiss}
\end{equation}

\subsection{Flow relation} Based on the reduced dissipation inequality \eqref{reddiss} we postulate the existence of a yield function $f$, which is a function of the generalized stress ${\sf S}$, and a flow relation that takes the form of a normality law: that is,
\begin{subequations}
\begin{align}
& f({\sf S} ) \leq 0 \,, \label{yield} \\
&\dot{\sf \Gamma} = \lambda \frac{\partial f}{\partial {\sf S}}\,, \\
& \lambda \geq 0,\quad f \leq 0,\quad \lambda f = 0 \,.
\label{flow1} 
\end{align}
\label{floweqns}
\end{subequations}
\begin{figure}[!ht]
 \centering
 \includegraphics[width = 0.8\textwidth]{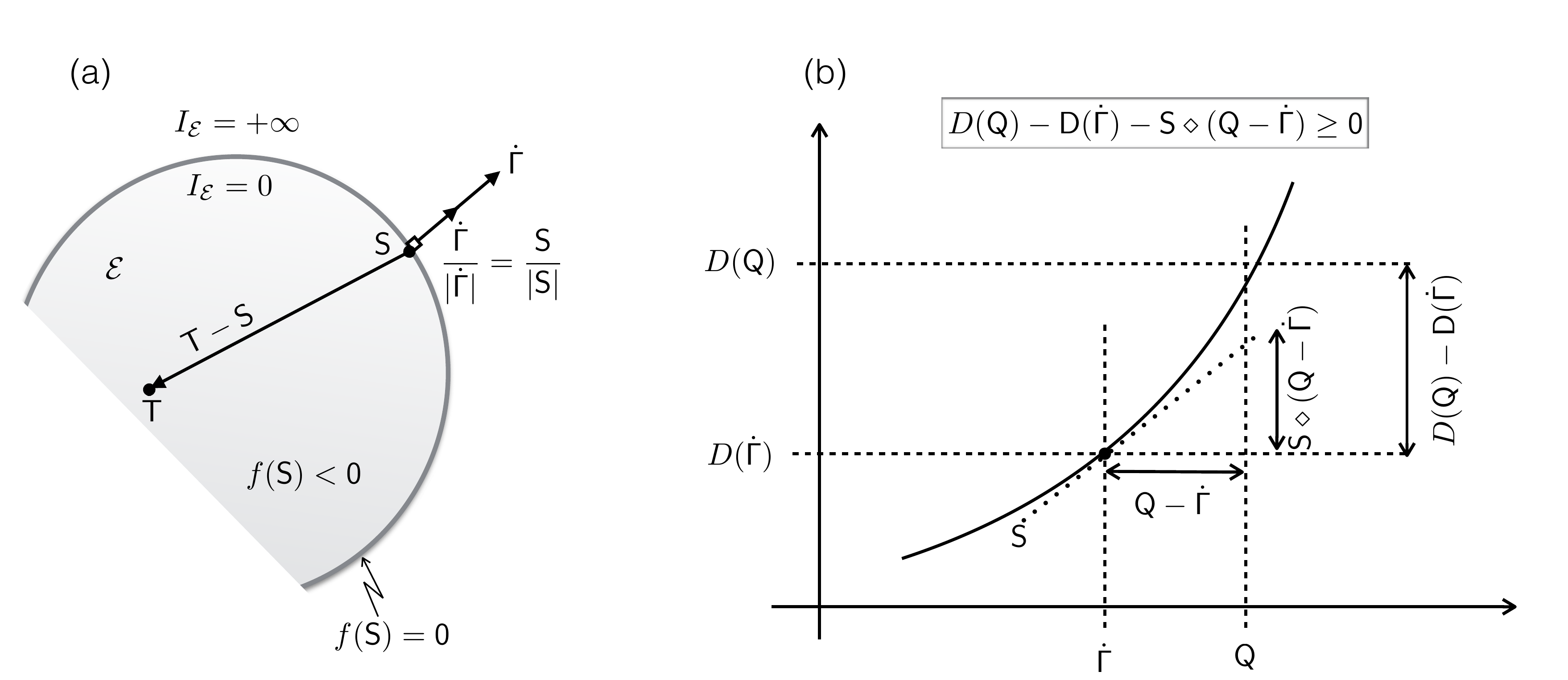}\\
 \caption{The yield surface and normality relation in generalized stress space 
 }
 \label{general_pics}
\end{figure}
Equivalently, as shown schematically in Figure \ref{general_pics}(a),
\begin{align}
&\dot{\sf \Gamma}\diamond ({\sf T} - {\sf S}) \leq 0 \qquad \mbox{for all}\ {\sf T} \in {\cal E}:= \{ {\sf T}\ |\ f({\sf T}) \leq 0 \},
\label{flow2}
\end{align} 
where ${\cal E}$ is the convex elastic region. 

The dissipation function $D$ may be defined using a generalization of the postulate of maximum plastic work in the form
\begin{equation}
D (\dot{\sf{\Gamma}}) = \sup \{ {\sf S\diamond\dot\Gamma}\ |\ f({\sf S}) \leq 0 \}\,.
\label{Ddef}
\end{equation}
Note that $D$ is convex and positively homogeneous, the latter being defined as $D(\alpha {\sf \dot{\Gamma}}) = |\alpha|D({\sf \dot{\Gamma}})$ for any real number $\alpha$.

{\bf Example}\ \ 
For the special but important case in which 
\begin{equation}
f({\sf S}) = |{\sf S}| - Y = \sqrt{|\bpi|^2 + \ell^{-2}|\bPi|^2} -Y \leq 0,
\label{f}
\end{equation}
where $Y$ is the yield stress, it follows from (\ref{floweqns}b) that at yield ($f = 0$)
\[
\lambda = |\dot{\sf \Gamma}| = \sqrt{|\dot\bepsilon^p|^2 + \ell^2|\nabla\dot\bepsilon^p|^2}\,.
\]
Furthermore, for non-zero ${\sf \dot\Gamma}$,
\begin{equation}
\frac{\dot{\sf \Gamma}}{|\dot{\sf \Gamma}|} = \frac{{\sf S}}{|{\sf S}|}\quad\Longleftrightarrow\quad {\sf S} = Y\frac{\dot{\sf \Gamma}}{|\dot{\sf \Gamma}|}\,.
\label{flowalt}
\end{equation}
From \eqref{Ddef} it is easily seen that for this example
\begin{equation}
D({\sf \dot\Gamma}) = Y|\dot{\sf \Gamma}|\,.
\label{Dexample}
\end{equation}
\makebox[1cm]{ } \hfill $\Box$

\begin{figure}[!ht]
 \centering
 \includegraphics[width = 0.6\textwidth]{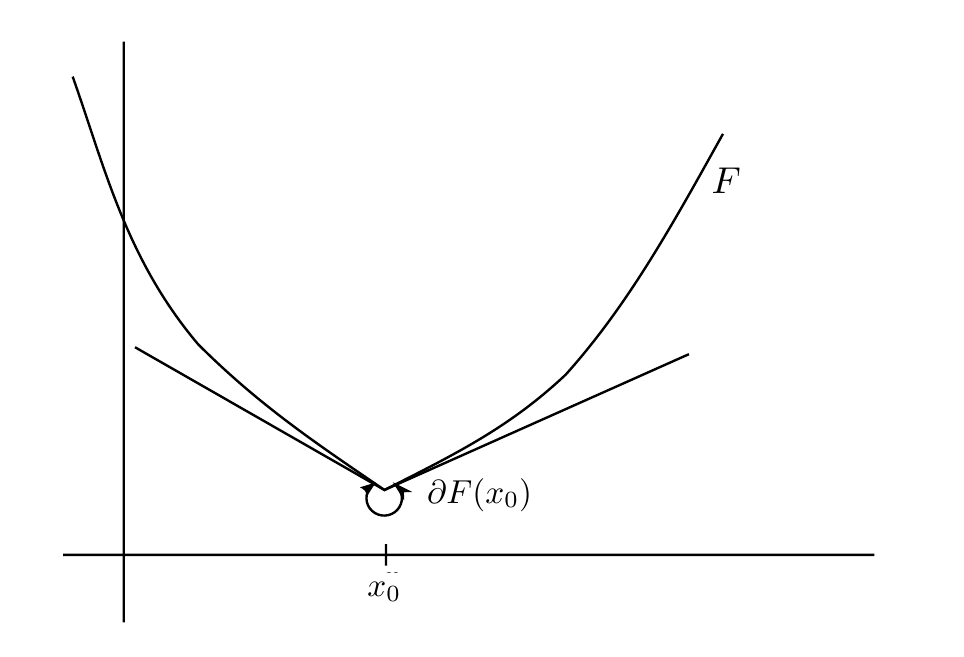}
 \caption{The subdifferential $\partial F(x_0)$ of a convex function $F$ at $x_0$}
 \label{subdifffig}
\end{figure}

There is an important duality between the flow relation written in terms of the yield and dissipation functions. To define this we need the notion of the subdifferential $\partial F$ of a convex function $F$, defined here on a finite-dimensional space $X$ such as $\BBR^d$: this is a set defined 
by\footnote{For this and other concepts from convex analysis, see for example \cite{HR}}
\begin{equation}
\partial F (\bx) = \{ \bp \ |\ F(\by) - F(\bx) - \bp\cdot(\by - \bx) \geq 0,\ \mbox{for all}\ \by \in X \}\,. 
\label{subdiff}
\end{equation}
That is, $\partial F$ is the set of tangents at the point $\bx$ (Figure \ref{subdifffig}). If $F$ is smooth at $\bx$ then $\partial F$ comprises a single member, viz. the tangent $\nabla F(\bx)$ to $F$ at $\bx$, or equivalently the gradient or normal to the level set $F = \mbox{constant}.$ 
Returning to plasticity, we define the indicator function $I_{\cal E}$ of a set (in this case the elastic region \eqref{flow2}) by
\begin{equation}
I_{\cal E}({\sf S}) = \left\{ \begin{array}{rl} 0 & \mbox{if}\ {\sf S} \in {\cal E}\,, \\
+\infty & \mbox{otherwise}\,.
\end{array} \right.
\label{indicator}
\end{equation}
This is a convex function. Furthermore, from the definition \eqref{subdiff} the subdifferential of $I_{\cal E}$ reads
\begin{equation}
\partial I_{\cal E} ({\sf S}) = 
\{ {\sf \dot\Gamma}\ |\ {\sf \dot\Gamma}\diamond ({\sf T} - {\sf S}) \leq 0\ \mbox{for all}\ {\sf T} \in {\cal E} \}\,.
\label{flow22}
\end{equation}
When compared with \eqref{flow2} we see that this is simply the \emph{normality relation}, albeit valid for a nonsmooth yield function. We use the notation 
\begin{equation}
N_{\cal E} ({\sf S})\quad \mbox{for}\quad \partial I_{\cal E} ({\sf S}) \,,
\label{normalcone}
\end{equation}
given its geometrical interpretation, and refer to $N_{\cal E}$ as the normal cone to ${\cal E}$ at ${\sf S}$. From the definition $N_{\cal E} = \{ {\sf 0} \}$ if ${\sf S}$ lies in the interior (that is, the elastic domain) of ${\cal E}$: as expected, the generalized plastic strain rate is zero if the generalized stress lies inside the elastic region.

From an important result in convex analysis we have the duality relation
\begin{equation}
{\sf \dot{\Gamma}} \in N_{\cal E} ({\sf S}) \quad \Longleftrightarrow\quad {\sf S} \in \partial D(\dot{\sf \Gamma}). 
\label{dualreln}
\end{equation}
The left-hand form of the normality relation has already been established. The equivalence \eqref{dualreln} indicates that it may also be written as 
\begin{equation}
D ({\sf Q} ) - D(\dot{\sf \Gamma}) - {\sf S}\diamond ({\sf Q - \dot\Gamma} ) \geq 0 \,,
\label{flow3}
\end{equation}
as depicted in Figure \ref{general_pics}(b). 
If $D$ is differentiable at $\dot{\sf \Gamma}$ then \eqref{flow3} reduces to the equation
\begin{equation}
{\sf S} = \left.\frac{\partial D}{\partial {\sf Q}}\right|_{\sf Q = \dot{\Gamma}}\,
\label{flowD}
\end{equation}
(replace ${\sf Q}$ by $\dot{\sf \Gamma} \pm \epsilon {\sf Q}$ and take the limit $\epsilon \rightarrow 0$). 

Two important examples of dissipation functions are 
\begin{equation}
D_1({\sf \dot\Gamma}) := Y\big[ |\dot\bepsilon^p| + \ell |\nabla\dot\bepsilon^p|\big]
\label{D1}
\end{equation} 
and
\begin{equation}
D_2({\sf \dot\Gamma}) := Y|{\sf \dot\Gamma}| = Y\sqrt{|\dot\bepsilon^p|^2 + \ell^2 |\nabla\dot\bepsilon^p|^2}\,.
\label{D2}
\end{equation}
\begin{figure}[!h]
\centering
\includegraphics[width=0.6\textwidth]{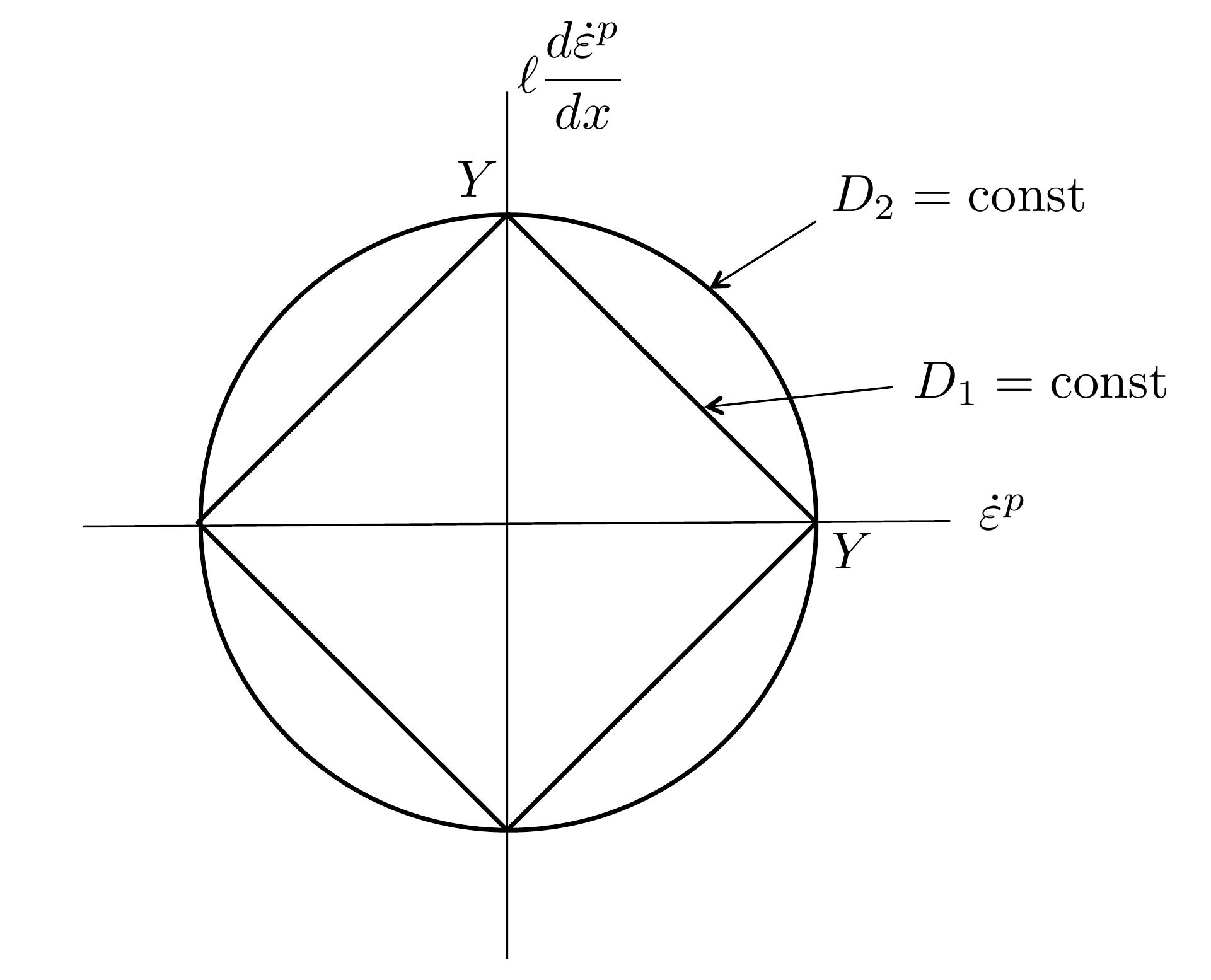}
\caption{The level sets corresponding to the dissipation functions $D_1$ and $D_2$}
\label{D1D2}
\end{figure}
The function $D_2$ corresponds to the definition \eqref{f} of the yield function. For $\dot{\sf \Gamma} \neq {\sf 0}$, from \eqref{flowD} with $D = D_2$ we recover the relation \eqref{flowalt}$_2$. Figure \ref{D1D2} shows the level sets corresponding to the dissipation functions $D_1$ and $D_2$. It is seen that  $D_2$ is smooth, and so is the corresponding yield function, while $D_1$ and its corresponding yield function are piecewise smooth. The dissipation function $D_1$ is of more than theoretical significance, in that Evans and Hutchinson \cite{EH} have shown theories based on such a dissipation to give results that correlate well with experiments on bending. The yield function corresponding to $D_1$ is shown by Reddy \cite{R} to be piecewise-smooth or Tresca-like in structure.

We now obtain a weak or global form for the flow relation with a view to eliminating ${\sf S}$ from it. Integrate (\ref{flow3}) to obtain
\begin{equation}
\int_\Omega \big[  D ({\sf Q} ) - D(\dot{\sf \Gamma}) - {\sf S}\diamond ({\sf Q - \dot\Gamma} )  \big]\ dx \geq 0 
\label{flow3int}
\end{equation}
and add to this the weak form of the microforce balance equation (\ref{mfbweak}) to get
\begin{equation}
\int_\Omega \big[ D ({\sf Q} ) - D(\dot{\sf \Gamma}) - \mbox{dev}\,\bsigma:(\bq - \dot\bepsilon^p)\big] \ dx\geq 0 \
\label{flow3int1}
\end{equation}
or
\begin{equation}
\int_\Omega \big[ D ({\sf Q} ) - D(\dot{\sf \Gamma}) - {\sf \Sigma}\diamond ({\sf Q - \dot\Gamma} ) \big]\ dx \geq 0 \
\label{flow3int2}
\end{equation}
where 
\[
{\sf \Sigma} := (\mbox{dev}\,\bsigma,\bzero )\,.
\]
Concepts such as the subdifferential defined earlier for vectors or tensors at a point (in essence, defined on $\BBR^d$) have a broader definition that extends to functionals. Thus, if we define the functional
\begin{equation}
j({\sf \dot\Gamma}) = \int_\Omega D({\sf \dot\Gamma})\ dx,
\label{j}
\end{equation}
then the subdifferential of $j$ at ${\sf \dot\Gamma}$ is defined to be the set of functions
\begin{equation}
\partial j({\sf \dot\Gamma}) = \Big\{ {\sf \Sigma} |\ j({\sf Q}) - j({\sf \dot\Gamma}) - \int_\Omega {\sf \Sigma}\diamond({\sf Q - \dot\Gamma})\ dx \geq 0\Big\}\,. 
\label{subdiffj}
\end{equation}
So we see that \eqref{flow3int2} corresponds to the \emph{global} statement that 
\begin{equation}
{\sf \Sigma} \in \partial j (\dot{\sf \Gamma})\,.
\label{Sinj}
\end{equation}
Furthermore, as in the local case the dual of this relation gives a global normality relation, which we write as 
\begin{equation}
\dot{\sf \Gamma} \in {\cal N}_{\cal E{\rm glob}}({\sf \Sigma})\,.
\label{jstar}
\end{equation}
The relation \eqref{jstar} is equivalent to finding the global form of the normality relation and the corresponding yield function. This is not a trivial task, as we shall see
in Section \ref{finddual} where, even for a discrete and therefore finite-dimensional approximation to the problem, at best it is possible to find only an upper bound to the yield function.

{\bf Remark}
Note that the microstress ${\sf S}$ has been eliminated from the global flow relation. This will be important in interpreting the flow relation for the gradient problem, as the local form \eqref{floweqns} involves ${\sf S}$, which is {\em indeterminate} in the elastic region.

\subsection{A mixed formulation for the dissipation function $D_1$}
If the dissipation function $D$ were a function of two independent variables, it would be feasible to obtain the corresponding yield function and normality law \eqref{jstar} by appealing to standard results from convex analysis. The arguments in $D$ are however the plastic strain and its gradient, and this relationship between the two variables complicates the task of finding the yield condition. With this in mind we explore a mixed approach in which the plastic strain gradient is treated as an independent variable.

For convenience we make use of the dissipation function $D_1$ defined in (\ref{D1}), and introduce the auxiliary variable $\bP$, a third-order tensor defined by
\begin{equation}
\bP = \ell \nabla \bepsilon^p\,.
\label{E}
\end{equation}
The dissipation function is now a function of two independent variables and can be written
\begin{align}
D_1(\bepsilon^p,\bP) & := D_{10}(\bepsilon^p) + D_{01}(\bP)  \nonumber \\
& = Y |\bepsilon^p| + Y |\bP|\,.
\label{Dep}
\end{align}
The corresponding flow relation reads 
\begin{equation}
(\bpi,\bPi) \in \partial D_1(\dot{\bepsilon}^p,\dot{\bP})\,.
\label{flowmixed}
\end{equation}
Since the two arguments of $D_1$ are now independent we may use an identity (\cite{ET}, (Ch. III, eqn (4.17), page 61)) to obtain
\begin{align}
I_{\cal E} (\bpi,\bPi) 
& = I_{{\cal E}_{10}}(\bpi) + I_{{\cal E}_{01}}(\bPi)\,.
\label{Depstar}
\end{align}
Here 
$I_{\cal E}$ is the indicator function for the set ${\cal E}$, ${\cal E}_{10} = \{ \bpi\ |\ |\bpi | \leq Y\}$ and ${\cal E}_{01} = \{ \bPi\ |\ |\bPi | \leq Y\}$.
Thus the use of a mixed approach allows the corresponding elastic region to be obtained easily. 

%
The flow relation (\ref{flowmixed}) is
\begin{equation}
\int_\Omega D_1(\bq,\bQ)\ dx - \int_\Omega D_1(\dot{\bepsilon}^p,\dot{\bP})\ dx - \int_\Omega \big[
\bpi:(\bq - \dot{\bepsilon}^p) + \bPi\circ (\bQ - \dot{\bP})\big]\ dx \geq 0\,,
\label{flowmixedglobal}
\end{equation}
where $\bq$ and $\bQ$ are respectively an arbitrary plastic strain and auxiliary variable.
Set $\bq = \bq - \dot{\bepsilon}^p$ in (\ref{mfbweak}) and add to (\ref{flowmixedglobal}) to obtain
\begin{equation}
\int_\Omega D_1(\bq,\bQ)\ dx - \int_\Omega D_1(\dot{\bepsilon}^p,\dot{\bP})\ dx - \int_\Omega \big[
 \bPi\circ [(\nabla\bq - \bQ ) - (\nabla\dot{\bepsilon}^p - \dot{\bP})\big]\ dx - \int_\Omega \mbox{dev}\,\bsigma:(\bq - \dot\bepsilon^p)\ dx \geq 0\,.
\label{flowmixedglobal2}
\end{equation}
By setting first $\bQ$, and then $\bq$, equal to zero, we extract the two variational inequalities
\begin{subequations}
\begin{align}
\int_\Omega Y|\bq|\ dx - \int_\Omega Y|\dot{\bepsilon}^p|\ dx - \int_\Omega 
 \bPi\circ \nabla (\bq - \dot{\bepsilon}^p) \ dx - \int_\Omega \mbox{dev}\,\bsigma:(\bq - \dot{\bepsilon}^p)\ dx \geq 0\,, \\
\int_\Omega Y|\bQ|\ dx - \int_\Omega Y|\dot{\bP}|\ dx + \int_\Omega 
 \bPi\circ (\bQ - \dot{\bP})\ dx  \geq 0\, .
\end{align}
\label{2VIs}
\end{subequations}
To these must be added the weak form of (\ref{E}), that is,
\begin{equation}
\int_\Omega \bP\circ \bQ\ dx - \int_\Omega \ell\nabla\bepsilon^p \circ\bQ\ dx = 0 \quad \mbox{for all}\ \bQ\,,
\label{weakE}
\end{equation}
and the weak form of the equilibrium equation \eqref{equm} together with the boundary conditions \eqref{macrobcs}: that is, 
\begin{equation}
\int_\Omega \bsigma (\bu,\bepsilon^p):\bepsilon (\bv)\ dx = \int_\Omega \bb\cdot\bv\ dx + \int_{\partial \Omega_t} \bar{\bt}\cdot\bv\ ds\,,
\label{weakequil}
\end{equation}
in which the test functions $\bv$  satisfy the homogeneous boundary condition $\bv = \bzero$ on $\partial \Omega_u$. We omit details of the (standard) function space setting for the set of weak equations.

Equations (\ref{2VIs}), (\ref{weakE}) and \eqref{weakequil} constitute a mixed problem for $\bu, \bepsilon^p,\,\bP$ and $\bPi$. This appears to be a nonstandard mixed problem.

\section{The regularized problem}
\label{RefRegProblem}
Later, when developing a computational approach we will focus on the dissipation function $D_2$, which is an elliptical cone and therefore smooth everywhere except at the origin. It will be convenient to replace $D_2$ by a regularized approximation $D_{2\eta}$, defined for $\eta > 0$ by 
\begin{equation}
D_{2\eta}({\sf \Gamma}) = Y\sqrt{|\bepsilon^p|^2 + \ell^2 |\nabla\bepsilon^p|^2 + \eta^2} \,.
\label{D2eta}
\end{equation}
\begin{figure}[!h]
\centering
\includegraphics[width=0.6\textwidth]{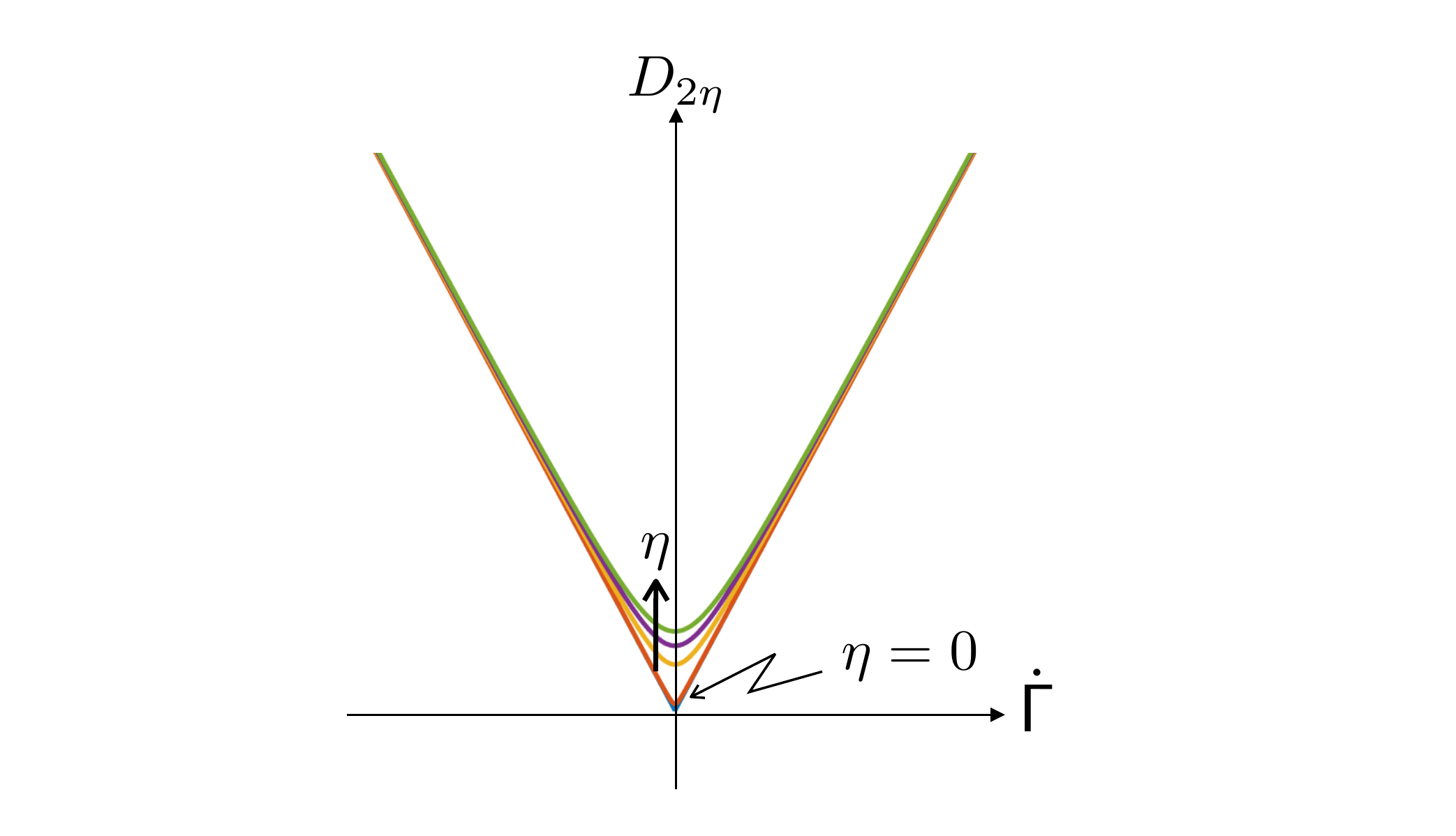}
\caption{The regularized dissipation function $D_{2\eta}$ }
\label{D2reg}
\end{figure}

The function is shown in Figure \ref{D2reg}. The local flow relation corresponding to $D_{2\eta}$ becomes, with the use of \eqref{flowD}, 
\begin{equation}
{\sf S} = \nabla D_{2\eta}(\dot{\sf \Gamma})\quad \Longleftrightarrow\quad \left\{ \begin{array}{l}
\bpi  = \displaystyle \left.\frac{\partial D_{2\eta}}{\partial \bq}\right|_{q = \dot{\varepsilon}^p} = \frac{Y^2\dot{\bepsilon}^p}{D_{2\eta}}     \,, \vspace{2ex} \\
\bPi = \displaystyle \left.\frac{\partial D_{2\eta}}{\ell \partial \nabla\bq}\right|_{\nabla q = \nabla \dot{\varepsilon}^p} = 
\frac{Y^2\nabla\dot\bepsilon^p}{D_{2\eta} }  \,.
\end{array} \right.
\label{flowlocalreg}
\end{equation}
Moreover, the inequality (\ref{flow3int}) becomes the equation
\begin{equation}
\int_\Omega \big[ \nabla D_{2\eta}(\dot{\sf \Gamma}) - {\sf \Sigma}\big]\diamond{\sf Q}\ dx = 0\,,
\label{weakD2eta}
\end{equation}
or
\begin{equation}
\int_\Omega  \left[ \Big(\frac{Y^2}{D_{2\eta}}\dot{\bepsilon}^p - \mbox{dev}\,\bsigma\Big):\bq 
+ 
\frac{Y^2}{D_{2\eta}}\ell^2\nabla\dot{\bepsilon}^p\circ\nabla\bq\right]  \ dx = 0\,.
\label{weakD2eta3}
\end{equation}
Assuming sufficient smoothness, integrating by parts and making use of the boundary conditions \eqref{microbcs}, we obtain the weak equation
\begin{equation}
\int_\Omega  \Big[\Big(\frac{Y^2}{D_{2\eta}}\dot{\bepsilon}^p - \mbox{dev}\,\bsigma\Big) 
- \ell^2
\mbox{div}\,\Big(\frac{Y^2}{D_{2\eta}}\nabla\dot{\bepsilon}^p \Big)\Big]:\bq \ dx = 0\,.
\label{weakD2eta4}
\end{equation}
This leads to the pointwise relation
\begin{equation}
\mbox{dev}\,\bsigma = \frac{Y^2}{D_{2\eta}}\dot{\bepsilon}^p - \ell^2\div\left(\frac{Y^2 \nabla\dot{\bepsilon}^p}{D_{2\eta}}\right)\,.
\label{weakD2eta5}
\end{equation}
We note that $(\mbox{div}\,\nabla\dot\bepsilon^p)_{ij} = \dot{\varepsilon}^p_{ij,kk}$ and that this quantity is deviatoric if $\dot\bepsilon^p$ is.
Equation \eqref{weakD2eta5}, which will form the basis of the computational investigation reported in Section \ref{numerics},
could have been obtained directly by substituting the regularized version of the flow relation \eqref{flowlocalreg}
in the microforce balance equation \eqref{mfb}. Indeed, the first and second terms on the right-hand side of \eqref{weakD2eta5} correspond respectively to $\bpi$ and $-\mbox{div}\,\bPi$.

\section{A time-discrete minimization problem}
\label{RefMinProblem}
The global problem \eqref{flow3int2} does not have an equivalent minimization problem. However, the corresponding time-discrete problem may be posed as a minimization problem. We discretize in time by partitioning the time interval $[0,T]$ as $0 = t_1 < t_2 < \cdots < t_n < \cdots t_N = T$, set $w_n := w(t_n)$ and $\Delta w = w_{n+1} - w_n$ for any function $w$, and replace the time derivative $\dot{w}$ by its backward Euler approximation $\Delta w/\Delta t$. Then (\ref{flow3int2}) becomes
\begin{equation}
\int_\Omega \big[ D ({\sf Q} ) - D(\Delta{\sf \Gamma}) - {\sf \Sigma}_{n+1}\diamond ({\sf Q} - \Delta {\sf \Gamma} ) \big]\ dx \geq 0 \,.
\label{flow3intinc}
\end{equation}
Here we have mutliplied throughout by $\Delta t$, made use of the positive homogeneity of $D$, and replaced the arbitrary ${\sf Q}\Delta t$ by ${\sf Q}$.  Now from \eqref{Cdev}, and noting that $\Delta\bepsilon^p$ is deviatoric,
\begin{align}
\mbox{dev}\,\bsigma_{n+1} & = \mbox{dev}\,[\BBC (\bepsilon_{n+1} -  \bepsilon^p_n - \Delta\bepsilon^p )] 
\nonumber \\
& = \bsigma^{\rm tr} - 2\mu  \Delta\bepsilon^p ,
\label{sigman+1}
\end{align}
where 
\begin{align}
\bsigma^{\rm tr} & := \mbox{dev}\,[\BBC (\bepsilon_{n+1} - \bepsilon^p_n )] \nonumber \\
& = 2\mu (\mbox{dev}\,\bepsilon_{n+1} - \bepsilon^p_n )
\label{trialstress}
\end{align}
is a deviatoric trial elastic stress; that is, the deviatoric stress corresponding to purely elastic behaviour in the time step $t_n \rightarrow t_{n+1}$. Thus (\ref{flow3intinc}) becomes
\begin{equation}
\int_\Omega \big[ D ({\sf Q} ) - D(\Delta{\sf \Gamma}) + 2\mu\,\Delta\bepsilon^p:(\bq - \Delta \bepsilon^p)
- \bsigma^{\rm tr}:(\bq - \Delta \bepsilon^p)
 \big]\ dx \geq 0 \,.
\label{flow3intinc2}
\end{equation}
This is equivalent to the minimization problem
\begin{equation}
\Delta\bepsilon^p = \mbox{argmin}_{\sf Q} L({\sf Q}),
\label{minprob}
\end{equation}
where
\begin{equation}
L({\sf Q}) := \int_\Omega \big[ D({\sf Q}) + \mu\,|\bq|^2 - \bsigma^{\rm tr}:\bq \big]\ dx\,,
\label{J}
\end{equation}
for given $\bsigma^{\rm tr}$ where as before ${\sf Q} = (\bq,\ell\nabla \bq)$. Note that, unlike the classical case, this is a global problem which cannot be reduced to a local or pointwise one, given that ${\sf Q}$ involves $\bq$ and its gradient.

\section{The spatially discrete problem}
\subsection{Discrete flow relations}
In this section we examine features of the spatially discrete problem. We also discretize in time as in Section \ref{RefMinProblem}.
The domain, assumed polygonal (in two dimensions) or polyhedral (in three) for convenience, is covered by a mesh comprising 
\begin{equation}
{\rm NE}\ \mbox{elements}\ \ \mbox{and}\ \  {\rm NN}\ \mbox{nodes}
\label{nodes}
\end{equation}
where NN excludes those nodes at which the plastic strain is prescribed. The number of plastic strain degrees of freedom at each node is, taking into account the symmetry of $\bepsilon^p$ and the plastic incompressibility condition $\mbox{tr}\,\bepsilon^p = 0$, 
\begin{equation}
{\rm ndofs} = d(d+1)/2 - 1
\label{ndofs}
\end{equation}
for a $d$-dimensional problem ($d > 1$). 

Denote the global degrees of freedom of $\bepsilon^p$ by ${\sf p}$ and those of the displacement by ${\sf d}$, and assume conventional  conforming approximations. Then
\begin{equation}
\bepsilon^p = {\sf Np}\,,\quad \nabla\bepsilon^p = {\sf Bp}\,,\quad \bu = \overline{\sf N}{\sf d},\quad \bepsilon (\bu) = \overline{\sf B}{\sf d},
\label{findim}
\end{equation}
where ${\sf N}$ and $\overline{\sf N}$ are matrices of shape functions and ${\sf B}$ and $\overline{\sf B}$ matrices of shape function derivatives. 

Here and elsewhere we drop the subscript $n$ that denotes quantities at time $t_n$.

Since 
\[
|\bepsilon^p | = \sqrt{{\sf p}^T{\sf N^TN}{\sf p}}\qquad\mbox{and}\qquad |\nabla\bepsilon^p | = \sqrt{{\sf p}^T{\sf B^TB}{\sf p}},
\]
we have, from (\ref{D1}), 
\begin{equation}
D_{1}({\sf p}) =  Y \Big[  \sqrt{{\sf p}^T{\sf N^TN}{\sf p}} +  \ell\sqrt{{\sf p}^T{\sf B^TB}{\sf p}}\,     \Big] ,
\label{D1discr}
\end{equation}
which is homogeneous of degree 1 in ${\sf p}$. Likewise,
\begin{align}
D_2({\sf p}) & =  Y \Big[  \sqrt{{\sf p}^T{\sf N^TN}{\sf p} + \ell^2{\sf p}^T{\sf B^TB}{\sf p}}\,     \Big]  \nonumber \\
& = Y  \sqrt{{\sf p}^T{\sf K}{\sf p} },
\label{D2discr}
\end{align}
where the pointwise matrix ${\sf K}$ is defined by 
\[
{\sf K = N^TN + \ell^2B^TB}\,.
\]
Next, set
\begin{equation}
{\cal J}_i ({\sf q}) := \int_\Omega D_i({\sf q})\, dx\qquad (i=1,2);
\label{jdiscr}
\end{equation}
then (\ref{flow3int}) becomes, for the incremental problem, 
\begin{align}
& {\cal J}_i({\sf q}) - {\cal J}_i(\Delta{\sf p}) - ({\sf q} - \Delta{\sf p})^T {\sf s} \geq 0 \quad\mbox{for all}\ {\sf q},
\label{Sinjdicr}
\end{align}
where the global vector of nodal stresses ${\sf s}$ is given by
\begin{equation}
{\sf s} := \int_\Omega {\sf N^T}\mbox{dev}\,\bsigma\,dx\,.
\label{sdef}
\end{equation}
Thus we have the discrete inclusion
\begin{equation}
{\sf s} \in \partial {\cal J}_i(\Delta{\sf p})
\label{Asinj}
\end{equation}
and the dual of this is, from \eqref{dualreln}, 
\begin{align}
\Delta{\sf p} \in N_{{\cal E}_i}({\sf s})\quad\mbox{or}\quad (\Delta{\sf p})^T({\sf t - s}) \leq 0\quad\mbox{for all}\ {\sf t} \in {\cal E}_i\,,
\label{dualdiscr}
\end{align}
in which ${\cal E}_i$ is the elastic region, in the space of discrete stresses ${\sf s}$, corresponding to the dissipation function ${\cal J}_i$. 
As with the continuous problem the inclusion (\ref{Asinj}) is equivalent to a minimization problem. First, we have
\begin{equation}
{\sf s} = \underbrace{\int_\Omega {\sf N}^T\bsigma^{\rm tr}\ dx}_{{\sf s}^{\rm tr}} - \underbrace{\int_\Omega 2\mu\,{\sf N}^T{\sf N}\ dx}_{\sf M}\, \Delta {\sf p},
\label{decomps}
\end{equation}
so that the minimization problem is
\begin{equation}
\Delta {\sf p} = \mbox{argmin}_{\sf q} \ \big({\cal J}_i({\sf q}) + {\sf q}^T{\sf M}{\sf q} - {\sf q}^T{\sf s}^{\rm tr}\big)\,.
\label{minprobdiscr}
\end{equation}
Thus we have obtained a vehicle to establish the relation between the dissipation function corresponding to the global dissipation functions ${\cal J}_i$ and their corresponding elastic regions ${\cal E}_i$. 

\subsection{Finding the elastic region}\label{finddual}
\newcommand{\normp}{\|\hspace{-0.3ex}|}
We have available the global dissipation functions ${\cal J}_i$ and now seek to construct the corresponding elastic regions ${\cal E}_i$ and associated yield functions, which would allow the use of the flow law as a normality relation, as in \eqref{dualdiscr}. Now from a result in convex analysis (see for example \cite{HR}, page 109), given a dissipation function ${\cal J}$, one may construct a yield function $\phi ({\sf s})$ as a function of the global nodal stresses with the properties
\begin{subequations}
\begin{align}
& \phi\ \mbox{is positively homogeneous and convex}\,, \label{phiposhom} \\
& {\cal E} = \{ {\sf s}\ |\ \phi({\sf s}) \leq 1 \}\,, \label{canonyield}  \\
& \phi({\sf s}) = \sup_{{\sf q}\neq {\sf 0}}\frac{{\sf q^Ts}}{{\cal J}({\sf q})}\,.
\label{phiprops}
\end{align}
\end{subequations}
It follows that the yield function can be constructed if we are able to evaluate the supremum in \eqref{phiprops}.

Locally, the relationship \eqref{phiprops} is exemplified in the yield and dissipation functions \eqref{f} and \eqref{D2}. Unfortunately, determining $\phi$ in \eqref{phiprops} corresponding to the global functions ${\cal J}_i$ is not a simple task, as will be seen: the best that can be done is to obtain a function that is an upper bound for $\phi$. To see this, we focus on the dissipation function ${\cal J}_2$ and note that this can be written, for constant yield stress $Y$, as
\[
{\cal J}_2({\sf q}) = Y \int_\Omega |{\sf K}^{1/2}{\sf q}|\ dx\,.
\]
Taking $|\Omega | = 1$ for convenience we note that 
\begin{align}
{\sf s}^T{\sf q} & = \int_\Omega ([{\sf K}(\bx)]^{-1/2}{\sf s})^T([{\sf K}(\bx)]^{1/2}{\sf q})\ dx \nonumber \\
& \leq \int_\Omega |[{\sf K}(\bx)]^{-1/2}{\sf s}|\,|[{\sf K}(\bx)]^{1/2}{\sf q}|\ dx \nonumber \\
& \leq  Y^{-1}\big(\mbox{max}_{\bx \in \Omega} |[{\sf K}(\bx)]^{-1/2}{\sf s}|\big) {\cal J}({\sf q})\,. 
\label{Cauchy}
\end{align}
Hence we have, from \eqref{phiprops} and \eqref{Cauchy},
\begin{align}
\phi({\sf s}) & =  \sup_{{\sf q \neq 0}}\frac{\sf q^Ts}{\displaystyle Y\int_\Omega |{\sf K^{1/2}}(\bx){\sf q}|\ dx} \nonumber \\
& \leq 
Y^{-1}\mbox{max}_{\bx \in \Omega} |[{\sf K}(\bx)]^{-1/2}{\sf s}|\,.
\label{normest}
\end{align}
In order for the expression on the right-hand side of \eqref{normest} to be equivalent to the yield function $\phi$, the supremum in the first line of \eqref{normest} has to be achieved at this value. That is, assuming the supremum to be achieved for $\overline{\sf q} \neq {\sf 0}$, we must have
\begin{equation}
\frac{\sf \overline{q}^Ts}{\displaystyle \int_\Omega |{\sf K^{1/2}}(\bx){\sf \overline{q}}|\ dx} \nonumber \\
=
\mbox{max}_{\bx \in \Omega} |[{\sf K}(\bx)]^{-1/2}{\sf s}|\,,
\label{sup}
\end{equation} 
or
\begin{equation}
\frac{\displaystyle \int_\Omega   ([{\sf K}(\bx)]^{-1/2}{\sf s})^T([{\sf K}(\bx)]^{1/2}{\sf \overline{q}})\ dx}{\displaystyle \int_\Omega |{\sf K^{1/2}}(\bx){\sf \overline{q}}|\ dx} \nonumber \\
=
\mbox{max}_{\bx \in \Omega} |[{\sf K}(\bx)]^{-1/2}{\sf s}|\,.
\label{sup}
\end{equation} 
Since this equation must hold for any ${\sf s}$, we require that ${\sf K}$ be constant, which is a contradiction.

\section{Numerical investigation}\label{numerics}
We consider fully discrete approximations of the problem, based on weak forms of the equilibrium, microforce balance, and flow relations, with time-discretization as set out in Section \ref{RefMinProblem}, and making use of the regularized form \eqref{D2eta} of the dissipation function. Assume that the state of the system is known at time $t_n$ and that a backward Euler time-integration scheme is employed.
The weak form of the equilibrium and microforce balances (\ref{equm})--(\ref{mfb}) for the system at $t_{n+1} = t_n + \Delta t$ (the system of residual equations) are given by
\begin{align}
R_{\sf d} &:= \int_\Omega  \b{\varepsilon}(\b{v}) : \b{\sigma}_{n+1} \,dx 
- \int_{\partial \Omega_t} \b{v} \cdot \o{\b{t}}_{n+1} \,ds  \,  , \label{R_d}\\
R_{\sf p} &:=  \int_\Omega \b{q} : \text{dev}\, \b{\sigma}_{n+1} \, dx
-\int_\Omega  \b{q} : \b{\pi}_{n+1} \, dx
-\int_\Omega \nabla \b{q} : \b{\Pi}_{n+1} \, dx  \, , \label{R_p}
\end{align}
where as before $\bv$ and $\bq$ are respectively displacement and plastic strain test functions, 
 $\b{\pi}$ and $\b{\Pi}$ are given by \eqref{flowlocalreg}, and 
\begin{align*}
\b{\sigma}_{n+1} = \mathbb{C}(\b{\varepsilon}(\b{u}_{n+1}) - \b{\varepsilon}^\text{p}_{n+1}) .
\end{align*}
Equation \eqref{R_p} incorporates both microforce balance and the flow relation, and after discretization is therefore equivalent to the minimization problem \eqref{minprobdiscr}.

Denote the rate of change of an arbitrary quantity within a time-step by $\d (\bullet) := ((\bullet)_{n+1} - (\bullet)_{n}) / \Delta t$. For convenience we make use of the regularized version \eqref{D2eta} of the dissipation function. The problem then becomes one of solving a smooth set of equations. The magnitude of the perturbation $\eta$ is chosen to be small enough for trends in the elastic-plastic behaviour to be captured with sufficient accuracy. 


It has been shown in \cite{R} that a sufficient condition for the existence of a unique solution to the purely dissipative problem is that there be some hardening present. Accordingly, we introduce a small amount of hardening to avoid pathologies in the numerical solutions; the hardening may be viewed as a small perturbation, which does not affect the overall features of the solutions. 

Plastic incompressibility is enforced via the inclusion of the energy term
\begin{align*}
\psi_\text{inc} = \dfrac{\beta}{2} (\tr\, {\b{\varepsilon}}^p)^2 \, ,
\end{align*}
whose derivative with respect to the vector ${\sf p}$ is added to $R_{\sf p}$, where $\beta >0$ is a penalty.

The problem is then one of solving equations \eqref{R_d} and \eqref{R_p} for the displacement $\b{u}_{n+1}$ and plastic strain increment $\d\b{\varepsilon}^p$.


The displacement and plastic strain fields (and their associated test functions) are approximated using conforming $Q1$ interpolations.
The vector of global unknowns is denoted by $\sf{x} := ( \sf{d} , \sf{p})$.
A global Newton--Raphson procedure is used to linearize and iteratively solve the system of residual equations. 
An arbitrary variable evaluated at the current iteration $(i)$ in time step $n+1$ is denoted by $(\bullet)^{(i)}_{n+1} \equiv (\bullet)^{(i)}$.
The linearized problem and the iterative update of the solution vector are  given by
\begin{align*}
\dfracp{\sf R}{\sf x} \biggr|_{(i)} \cdot \varDelta {\sf x} = -\sf{R}^{(i)} \, ,\\
{\sf x}^{(i+1)} = {\sf x}^{(i)} +   \varDelta {\sf x} \, ,
\end{align*}
where ${\sf R}:= ( R_{\sf d}, R_{\sf p} )$. 

The finite element library AceGen \cite{ACE} is used to implement the finite element interpolation,  and to compute the residual and  tangent contributions  using automatic differentiation. 
This approach greatly simplifies the implementation. 
In addition, an adaptive time-stepping algorithm is employed.

We consider two examples, viz. a thin plate is subjected to a biaxial deformation, and uniaxial extension of a rod. Before doing so, and in order to contextualise those results, we  briefly review the study by  Fleck, Hutchinson and Willis \cite{FHW1,FHW2} of a strip in tension that is subjected to passivation on two surfaces at a certain point in its loading history.

\subsection{The problem of a strip in tension}
The problem is one in plane strain, of a strip $(-\infty\times \infty) \times (-h,h)$ that is subjected to a uniform applied strain $\varepsilon_{11} \equiv\bar{\varepsilon}$ in the $x$-direction. The only non-zero plastic strain components are $\varepsilon_{11}^p$ and $\varepsilon_{22}^p = -\varepsilon_{11}^p$, which follows from plastic incompressibility.

The surfaces $y = \pm h$ are initially traction-free and micro-traction free. At a certain point in the loading history beyond that of initial yield these surfaces are passivated, resulting in the plastic strain rate being zero on the boundaries from this point onwards. 
The authors in \cite{FHW1} report an \emph{elastic gap}: that is, purely elastic behaviour following passivation, with plastic flow occurring after the load has increased somewhat. 

The strain $\varepsilon_{11} = \bar{\varepsilon}$ is prescribed and increases monotonically. We therefore use the time $t$ as a parameter. 

\begin{figure}[h!]
\centering
\includegraphics[width=0.6\textwidth]{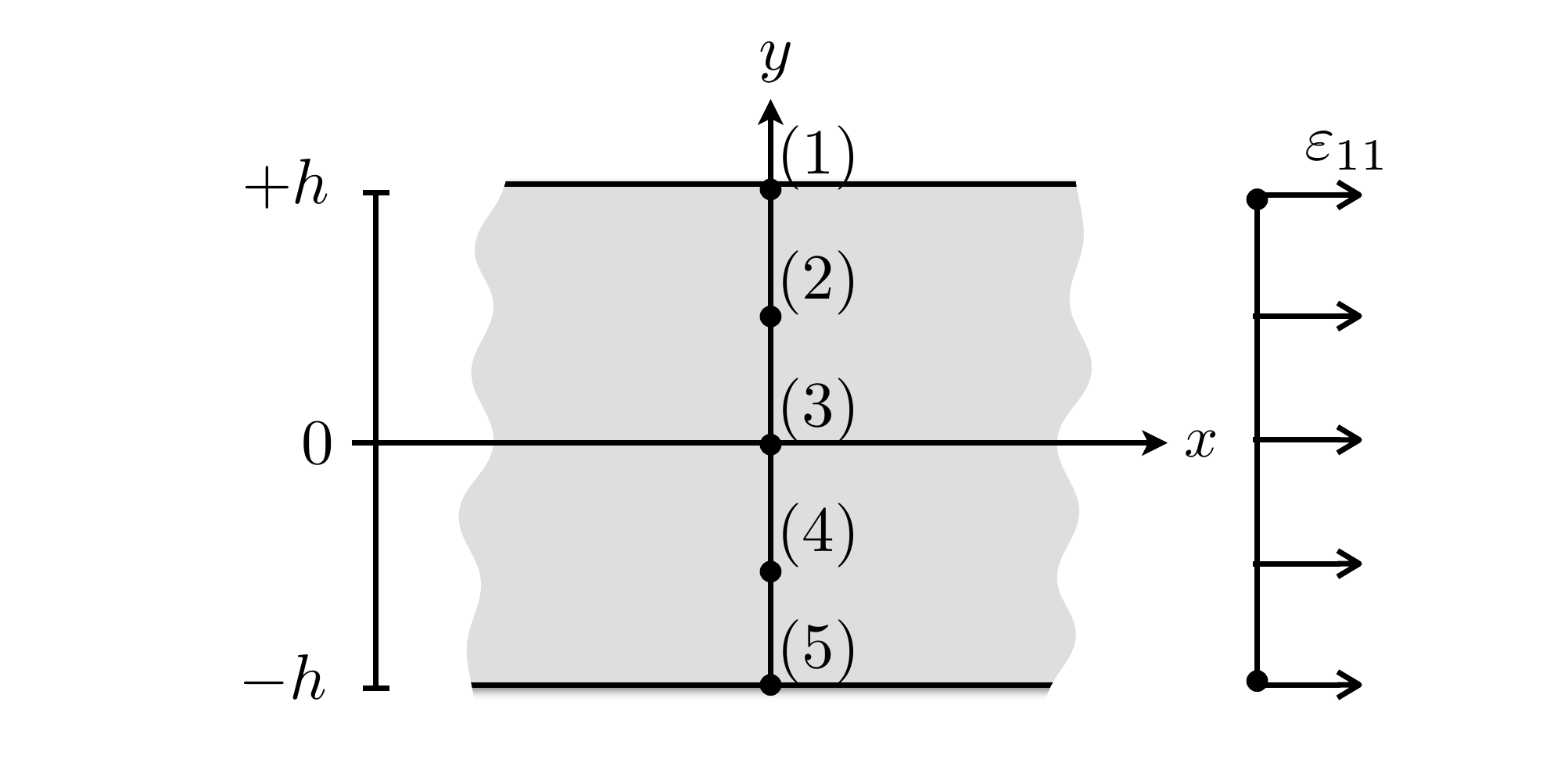}
\caption{Finite element mesh for the problem of a strip in uniform tension}
\label{strip_tension}
\end{figure}

The problem in question is one-dimensional, so for definiteness consider a mesh of uniform 1D elements with nodes 1 - 5 located respectively at $x = 0$ and $y = h, h/2, 0, -h/2, -h$ (Figure \ref{strip_tension}). From symmetry ${\sf p}_2 = {\sf p}_4$ and ${\sf p}_1 = {\sf p}_5$. 

{\em The uniform phase}\\
For time steps $t_1, t_2, \ldots, t_n$, ${\sf p}_1 = {\sf p}_5 \neq {\sf 0}$ and  
\begin{equation}
({\sf p}_i)_n \equiv \overline{{\sf p}}\quad (i=1,\ldots 5)\,.
\label{nonpass}
\end{equation}
\
Assume that the yield stress $Y$ is given. 
For the case of uniform deformation $\nabla \bepsilon^p = \bzero$ and the dissipation and yield functions are the conventional ones. 

%

{\em The passivation phase}\\
Consider the first time step following passivation: we now have $\Delta {\sf p}_1 = \Delta {\sf p}_5 = \bzero$ and there are only three free degrees of freedom, corresponding to nodes $2,\,3,\,4$. Each $\Delta {\sf p}_i$ has one independent component since there is no shear, and from plastic incompressibility $\varepsilon^p_{22} =-\varepsilon^p_{11}$.

Some insight into the elastic gap may be gained by making use of the definition \eqref{phiprops} of the canonical yield function $\phi$. Denote by $N_{\rm unif}$ the number of nodal degrees of freedom of plastic strain: this will be equal to the total number of nodes, since the boundary condition is micro-traction free. Likewise, denote by $N_{\rm pass}$ the number of degrees of freedom in the passivation phase. We have $N_{\rm pass} < N_{\rm unif}$ as the plastic strain increment is prescribed to be zero on the boundary nodes. Assuming for convenience that the difference between the  vectors ${\sf s}$ of nodal stresses just before and after initiation of passivation is negligible, it follows from \eqref{phiprops} that
\begin{align}
\phi_{\rm pass}({\sf s}) & =  \sup \frac{{\sf s^T}{\sf q}_{\rm pass}}{{\cal J}({\sf q}_{\rm pass})} \nonumber \\ 
& \leq \sup \frac{{\sf s^T}{\sf q}_{\rm unif}}{{\cal J}({\sf q}_{\rm unif})} \nonumber \\
& = \phi({\sf s}_{\rm unif}) = 1\,.
\label{unifpass}
\end{align}
Here we have denoted by ${\sf q}_{\rm unif}$ and ${\sf q}_{\rm pass}$ arbitrary vectors in the uniform and passivated phases, respectively. Also, in the last line we use the assumption that the material is in the plastic range in the uniform phase just before passivation. The inequality in the second line follows from the fact that ${\rm dim}\,{\sf q}_{\rm pass} = N_{\rm pass} < N_{\rm unif} = {\rm dim}\,{\sf q}_{\rm unif}$, so that the supremum is being taken over a larger set.
From this bound it is clearly possible that $\phi_{\rm pass} < 1$, so that the response {\em could be elastic} in the initial passivation phase.

When plastic flow does eventually take place, the inclusion (\ref{Asinj}) gives an explicit expression for the stress, viz.
\begin{align}
{\sf s}_{n+1} & = \left. \frac{\partial {\cal D}}{\partial {\sf p}}\right|_{\Delta {\sf p}} \nonumber \\
& = \int_\Omega \frac{{\sf K}\Delta {\sf p}}{\sqrt{(\Delta {\sf p})^T{\sf K}\Delta{\sf p}}}\ dx\,.
\label{flowreln}
\end{align}
Unlike the classical case, this cannot be inverted.

{\bf Remark}\ \  It is worth noting that, in \cite{FHW1}, the authors make use of the conventional Mises yield condition to determine yield, whereas we have shown that the condition is a more complex, global one. The assumption is also made in \cite{FHW1} that the stress state is uniform in the steps following passivation. This cannot be the case as the situation following passivation is non-uniform. 

\subsection{Biaxial deformation of a thin micro-plate}
In this example the role of the microscopic boundary conditions on the evolution of the problem are of particular interest. 
The material properties used in this example and the next are listed in Table~\ref{tab_material_props}, unless stated otherwise.

\begin{table}[htb]
\caption{Constitutive parameters used for the numerical examples unless stated otherwise}
\begin{center}
\begin{tabular}{l l l l}
First Lam\'{e} parameter 	& $\lambda$ 	& \num{1.05e-1} 	& \si{N/\micro\meter^2}   \\
Poisson's ratio	& $\nu$ 		& \num{0.3} 	&    \\
Initial slip resistance & $Y_0$		&  \num{1e-3}	& \si{N/\micro\meter^2} \\
Regularization parameter & $\eta$ & \num{5e-4} \\
Incompressibility penalty & $\beta$ & \num{1e6} & \si{N/\micro\meter^2}
\end{tabular}
\end{center}
\label{tab_material_props}
\end{table}
\begin{figure}[!ht]
 \centering
 \includegraphics[width = \textwidth]{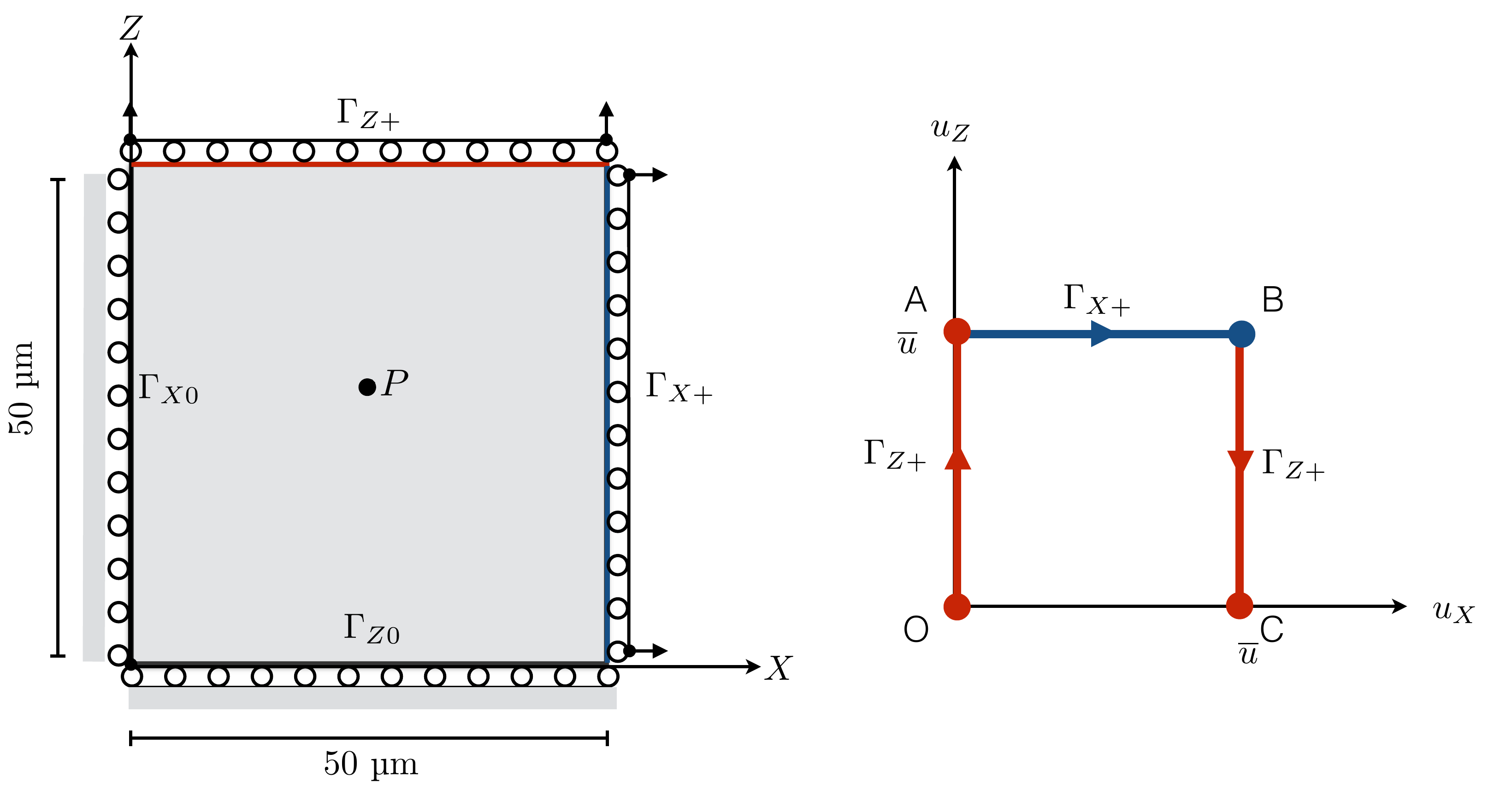}
 \caption{The problem of biaxial deformation in a plate and a schematic of the various stages of loading.}
 \label{bi_setup}
\end{figure}
Consider the $50 \times 2 \times 50$  \si{\micro\meter} plate shown in \fig{bi_setup}. 
In addition to the macroscopic boundary conditions indicated, the front face of the plate  ($y=0$) is prevented from displacing in the $y$-direction. 
The motion of the upper boundary $\Gamma_z^+$ and the right boundary $\Gamma_x^+$ is defined in three stages. 
During load stage O--A (see \fig{bi_setup}),  $\Gamma_z^+$ is displaced in the positive $z$-direction by a distance $\o{u} = 2/3$ \si{\micro\meter} while $\Gamma_x^+$ is prevented from displacing in the $x$-direction. 
During load stage A--B,  $\Gamma_x^+$ is displaced in the positive $x$-direction by a distance $\o{u}$ while $\Gamma_z^+$ is prevented from displacing in the $z$-direction. 
Finally, during stage B--C, the loading imposed during stage O--A is reversed by displacing $\Gamma_z^+$  a distance $-\o{u}$  in the $z$-direction, while $\Gamma_x^+$ is prevented from displacing in the $x$-direction.
Each loading stage (that is, OA, AB, BC and CA) corresponds to a time of $0.5$ \si{s}.
The plate is free to displace in the $y$-direction. 

The domain is discretized using $2500$ elements with one element through the thickness. 
The maximum permissible time-step size is \num{1e-3} \si{s}. 

The influence of a microscopic boundary condition for the plastic strain evolution on the global response is investigated by prescribing $\Gamma_z^+$  to be either (a) micro-free, (b) micro-hard from the onset, or (c) micro-free for $0\leq t < t_\text{pass}$, and thereafter preventing the evolution of plastic strain.
The boundary condition (c) is termed passivation. When imposed, passivation will occur at time $t_\text{pass} = 0.4~\si{s}$. 
The response of a material point in the centre of the specimen $P=[25,1,25]$ \si{\micro\meter} is monitored. 

\begin{figure}[!ht]
 \centering
 \includegraphics[width = \textwidth]{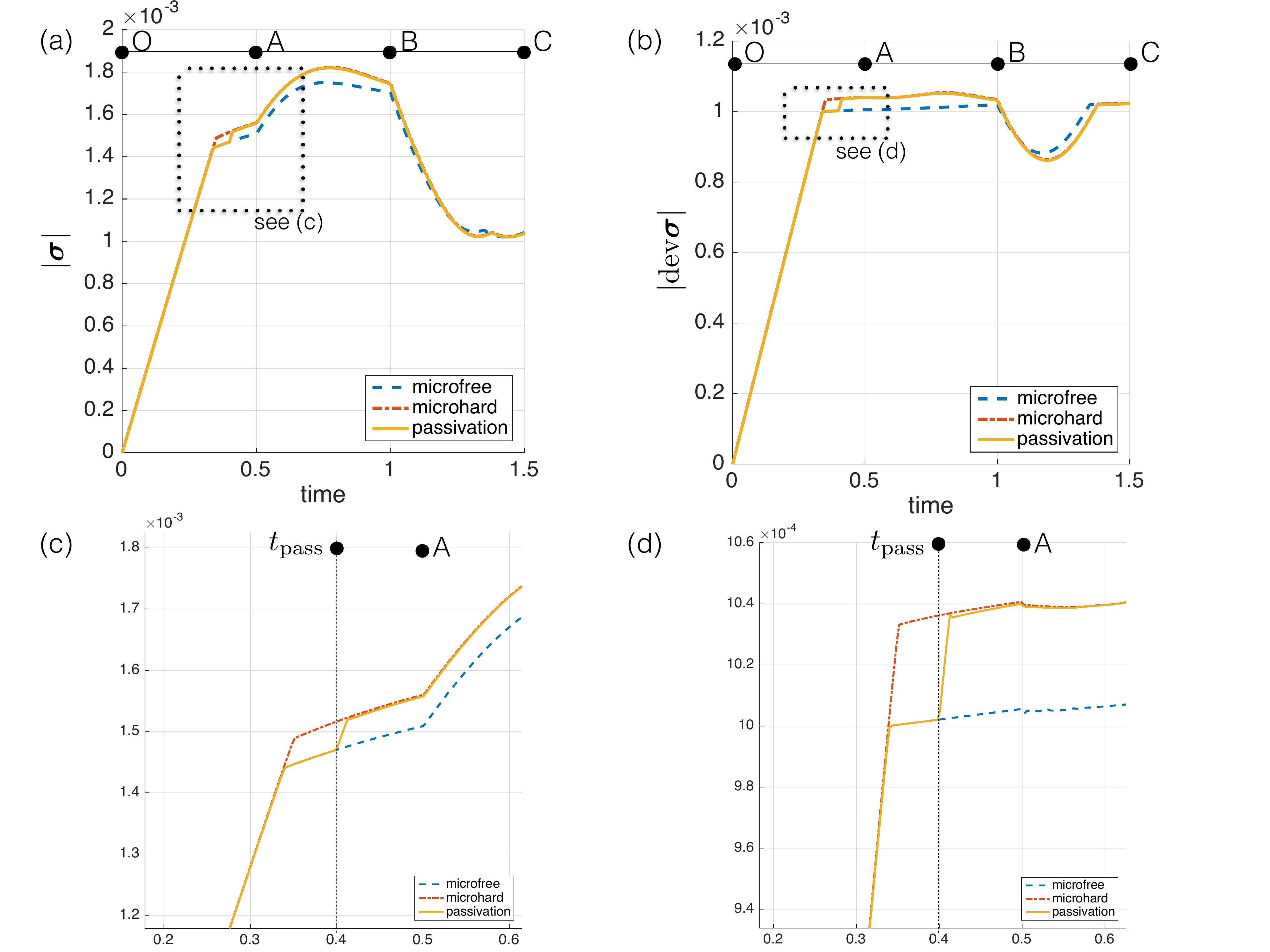}
 \caption{The evolution of (a) the Cauchy stress and (b) the Cauchy stress deviator at point P for the various boundary conditions. 
 The various load stages are also indicated.}
 \label{sigma_time_bi}
\end{figure}

The evolution of the magnitude of the Cauchy stress and stress deviator at the point $P$ for the various microscopic boundary conditions is shown in \fig{sigma_time_bi}. 
The macroscopic constraints give rise to the volumetric contributions to the stress tensor.
For the micro-free (a) and passivated (c) boundary conditions, yield occurs at $t := t_Y \approx 0.38$ \si{s} when $\vert \mbox{dev}\, \b{\sigma} \vert = Y_0$ (the initial yield stress). 
The onset of yielding at point $P$ is delayed when  micro-hard boundary conditions are imposed on the upper surface. 
Unsurprisingly, the presence of Dirichlet boundary conditions on the plastic strain changes the global response as the  residual expression (\ref{R_p})  now contains  additional constraints. 
As seen in  \fig{sigma_time_bi}(b) the amount of hardening is minimal.  
The hardening is linear for the micro-free problem but is more complex for the  micro-hard and passivated boundary conditions due to the contribution of the higher-order terms. 
Elastic unloading occurs at the onset of load stage B--C. 

The yield stress, that is, the stress at which global behaviour undergoes the transition from elastic to elastic-plastic, is determined automatically as a result of the perturbed dissipation function used in the computations. Thus this approach allows the yield stress to be obtained despite a closed-form expression not being available, as discussed in Section \ref{finddual}.

For the passivation problem (c), the microscopic Dirichlet constraints on the plastic strain evolution are imposed at $t_\text{pass} > t_Y$. 
The evolution of the stress state at point $P$ shows that the pre-passivation mico-free response transforms to a micro-hard response post-passivation. 
\emph{This transition occurs elastically. }
This is the phenomenon is referred to in \cite{FHW1} as an elastic gap. 
That the transformation occurs elastically is not surprising.
The system has to evolve to the curve corresponding to the micro-free boundary condition and it does so most efficiently, that is, by an elastic route. 

\begin{figure}[!ht]
 \centering
 \includegraphics[width = 0.9\textwidth]{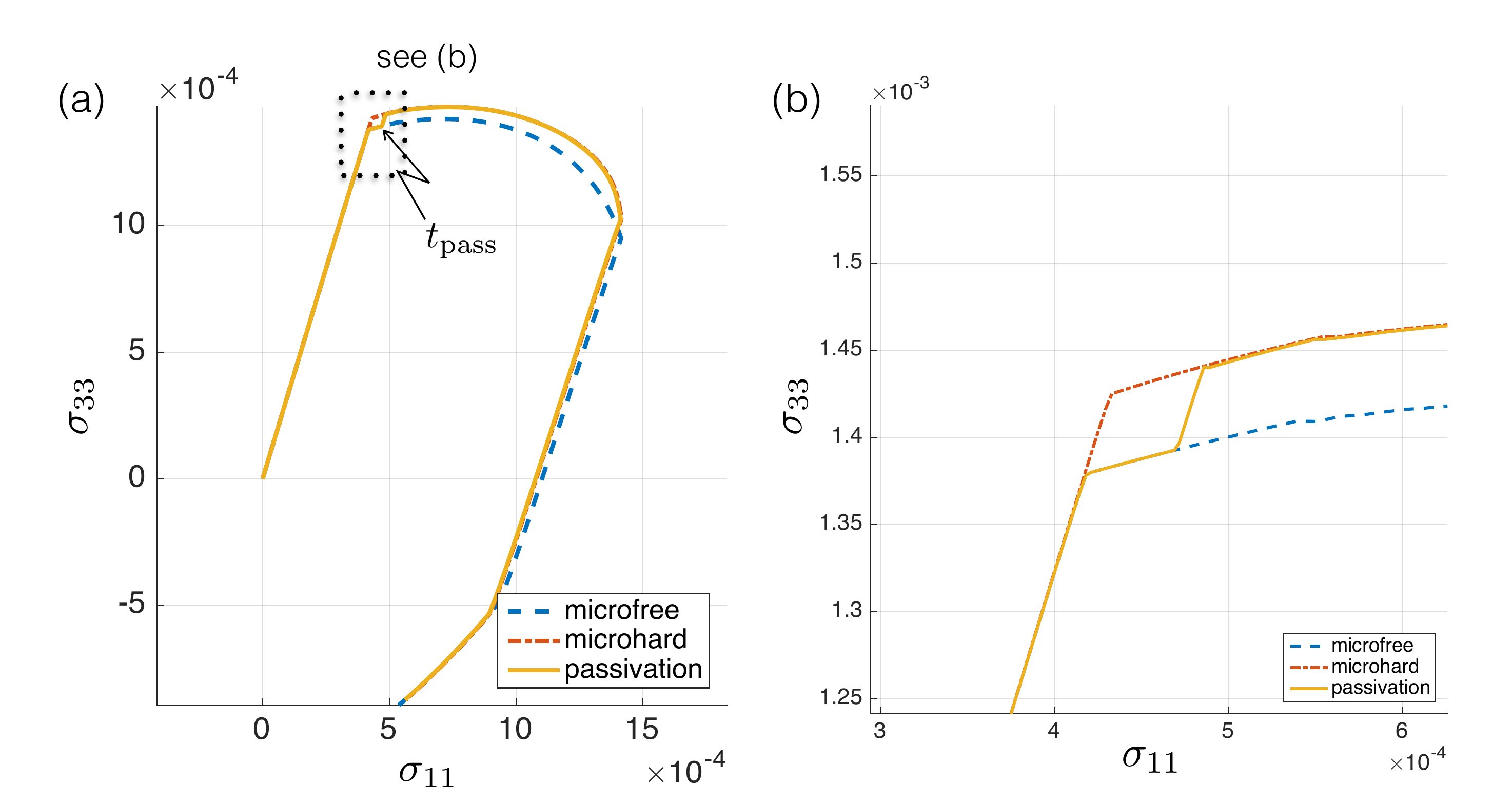}
 \caption{The evolution of the Cauchy stress components $\sigma_{11}$ and $\sigma_{33}$ for the various boundary conditions.}
 \label{yield_surface_bi}
\end{figure}

The form of the yield surface can be inferred from the stress response in the $\sigma_{11}$--$\sigma_{33}$ space, as shown in \fig{yield_surface_bi}.
Although not indicated, the micro-free response is identical to that obtained using a classical return mapping algorithm (closest-point projection) at the level of the integration point for the non-gradient, rate-independent $J_2$ plasticity problem. 
This confirms that the global formulation based on a micro-force balance is essentially equivalent to the local formulation in the absence of gradients. 
Formulations for the classical problem where the closest point projection occurs at the global level have been explored in \cite{SKT}. 
It also confirms that for the micro-free condition the choice of a primal formulation with the dissipation function $D_2$ is equivalent to the dual problem with a von Mises yield surface. 
The yield surface for the micro-free problem in \fig{yield_surface_bi} can thus be seen as the von Mises yield surface corresponding to the classical problem. 

The yield surface for the micro-hard problem is expanded relative to the micro-free one, consistently with the elastic gap transition reported earlier.
Furthermore, due to the gradient contributions to the hardening, the expansion is not uniform. 
The yield surface for the passivated problem is on the micro-free surface until $t=t_\text{pass}$, after which it moves elastically to the micro-hard one.
\emph{It should be noted that no elastic gaps occur with the onset of non-proportional loading corresponding to the transition from O--A to A--B. }

\subsection{Extension of a micro-rod}

\begin{figure}[!ht]
 \centering
 \includegraphics[width = 0.8\textwidth]{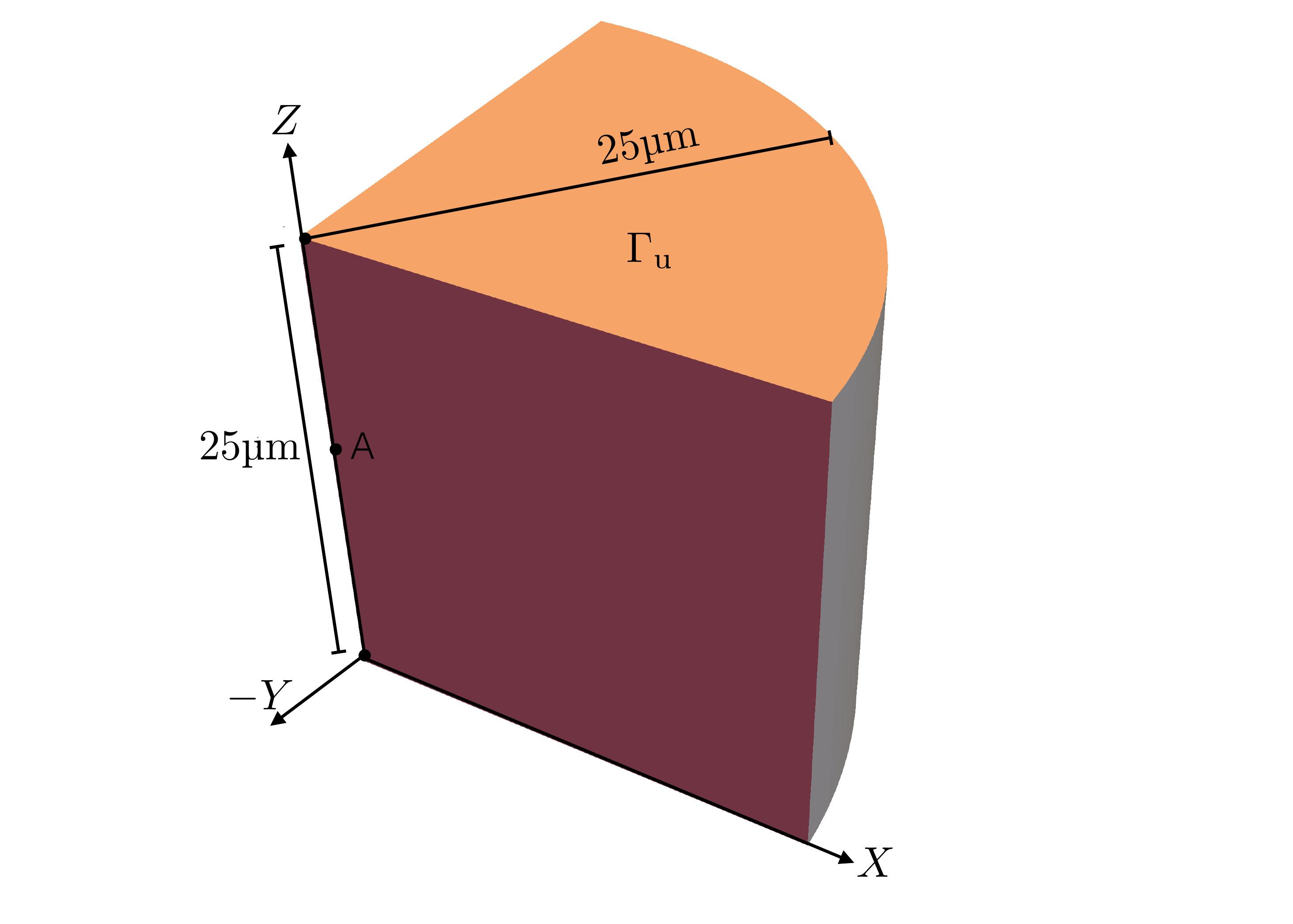}
 \caption{Computational domain for the problem of the extension of a rod.}
 \label{rod_setup}
\end{figure}

Consider a rod having radius 25 \si{\micro\meter} and length $L=50$ \si{\micro\meter}, and subjected to a prescribed displacement in the axial direction  of 0.5 \si{\micro\meter} applied to the upper and lower faces with normals $\b{e}_3$ and $-\b{e}_3$, respectively. 
Due to symmetry, only the upper quarter of the rod is modelled as shown in \fig{rod_setup}. 
The prescribed displacement is imposed incrementally over 0.5 \si{s}.
The response of the system at a material point labelled $A$ and located at $[0,0,12.5]$ \si{\micro\meter}  is recorded. 
The length scale, unless otherwise stated, is $l = 0.2 L$. 
The domain is discretized using 6527 elements. 

As in the previous example, the consequences of choosing different microscopic boundary conditions on the upper boundary of the domain, denoted $\Gamma_\text{u}$, are investigated. 
Passivation occurs at $t_\text{pass} = 0.25$ \si{s} which is well into the plastic range.

\begin{figure}[!ht]
 \centering
 \includegraphics[width = \textwidth]{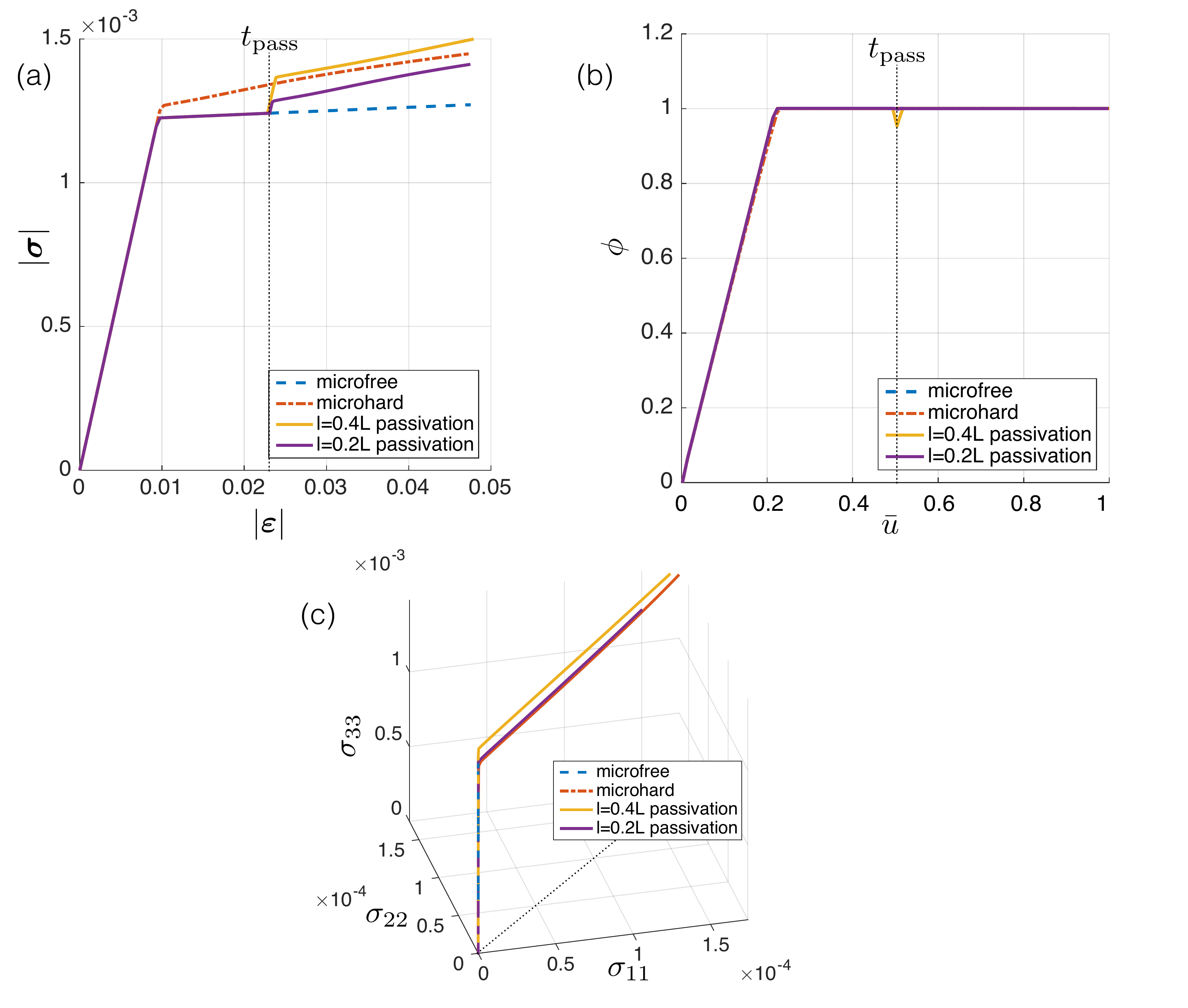}
 \caption{The evolution of the state at material point A for the rod extension problem for various microscopic boundary conditions and different length scales. 
 The relation between $\vert \b{\sigma} \vert$ and $\vert \b{\varepsilon} \vert$ is shown in (a). 
 The evolution of the canonical yield function $\phi$ and the Cauchy stress are shown in (b) and (c), respectively. The curve corresponding to the micro-hard boundary condition is for the length scale $\ell = 0.2L$.}
 \label{cylinder_results}
\end{figure}

The response at point A for the various choices of the microscopic boundary conditions on $\Gamma_\text{u}$ is shown in \fig{cylinder_results}. 
The relation between the magnitudes of the Cauchy stress and the strain,  shown in  \fig{cylinder_results}(a), clearly contain the same features discussed in the previous example:  an increase in the perceived yield strength for the micro-hard condition and an elastic gap for the passivation problem. 
Furthermore, the size of the elastic gap increases with increasing length scale. 
This relation between the size of the elastic gap and the length scale was also observed in \cite{FHW1}.

The evolution of the quantity $\phi := \vert  {\sf S} \vert / Y$, which corresponds to the classical yield function, is shown in \fig{cylinder_results}(b).
As expected, $\phi$ is in the range $0 <\phi < 1$ in the elastic region, and $\phi = 1$ during plastic flow for all microscopic boundary conditions. 
The elastic gap at $t=t_\text{pass}=0.25$ \si{s} is also clearly indicated for the passivation problems as $\phi$ drops below unity.

The evolution of the stress in the $\sigma_{11}$--$\sigma_{22}$--$\sigma_{33}$ space is shown in \fig{cylinder_results}(c).
For the micro-free condition, the stress state is uni-axial with $\sigma_{33}$ the only non-zero stress component.
The stress state remains at the point on the yield surface where initial yield occurred. 
The stress state is spatially uniform throughout the specimen and there are no plastic strain gradients present.

For the micro-hard boundary condition the stress evolves symmetrically in the $\sigma_{11}$ and $\sigma_{22}$ directions post yield. 
The micro-hard boundary condition constrains all components of the plastic strain, thereby inducing a stress response in directions other than that of the loading.
Microscopic Dirichlet conditions on the plastic strain result in plastic strain gradients and Cauchy stresses in directions other than the loading direction.  
The yield surface for the passivation problem is identical to that of the micro-free problem prior to passivation. 
The stress state at the point of passivation will have only a $\sigma_{33}$ component. 
Upon passivation, the elastic gap occurs. 
For the larger length scale of $l=0.4L$ the yield surface appears to increase above the micro-hard one. 
It should be noted that post-passivation, the stress components plotted are no longer the principal stress components.

\section{Concluding remarks}
A theoretical and computational investigation has been carried out of a dissipative model of rate-independent strain-gradient plasticity, that is, one in which gradient terms are accounted for only in the flow relation.  The global nature of the flow relation, previously reported in \cite{R}, is reiterated. The most appropriate and effective approach to formulating the flow relation is through the use of a dissipation function; this form of the relation is especially useful in the context of numerical investigations. Dual formulations in terms of the yield function and a normality relation have been approached using the tools of convex analysis. It is not possible, using conventional tools, to invert the flow relation to obtain the yield surface corresponding to the global dissipation function. This objective has been investigated further in the context of the fully discrete problem, for which an upper bound to the elastic region is found. 

The numerical investigation casts further light on the response using the dissipative theory in situations of non-proportional loading. Post-yield behaviour has been investigated. The elastic gap reported in \cite{FHW1} has been observed in situations in which passivation has been imposed. No such gap appears in cases of non-proportional loading in the form of a change in loading direction. It has been possible to interpret the gap mathematically, using the expression for the yield function as a maximum, taken over all admissible plastic strain increments, of a function involving the dissipation: the vector of admissible increments is necessarily smaller in dimension following passivation, and the corresponding maximum may therefore be smaller than that in the step preceding passivation. The elastic gap has also been observed to constitute an efficient ``transition" from a  stress-strain curve corresponding to a micro-free boundary condition, to that which is obtained assuming micro-hard boundary conditions. 

The dissipative model of strain-gradient plasticity has been shown in \cite{R} to be mathematically well posed. As has been indicated in \cite{FHW1}, experimental tests would clarify the predictive capabilities of this in relation to energetic strain-gradient models.

\section{Acknowledgements}
BDR acknowledges many helpful discussions on the topic of this paper with JW Hutchinson. The work reported in this paper was carried out with support through the South African Research Chair in Computational Mechanics to BDR and ATMcB. This support is gratefully acknowledged. 
PS acknowledges support through the Collaborative Research Center 814.


\begin{thebibliography}{}
\bibitem{Aifantis}
Aifantis, E.C. (1984). On the microstructural origin of certain inelastic models. J. Engng Mat.Tech. {\bf 106} 326--330.
\bibitem{ET} Ekeland, I. and Temam, R. (1976). Convex Analysis and Variational Problems. North-Holland, Amsterdam.
\bibitem{EH} Evans, A.G. and Hutchinson, J.W. (2009). A critical assessment of theories of strain gradient plasticity. Acta Materialia {\bf 57} 1675-1688.
\bibitem{FH} Fleck, N.A.  and Hutchinson, J.W. (2001). A reformulation of strain gradient plasticity. J. Mech. Phys. Solids {\bf 49}
2245--2271. 
\bibitem{FHW1} Fleck, N.A., Hutchinson, J.W. and Willis, J.R. (2015). Strain-gradient plasticity under non-proportional loading. Proc. R. Soc {\bf A 470} 20140267.
\bibitem{FHW2} Fleck, N.A., Hutchinson, J.W. and Willis, J.R. (2015). Guidelines for constructing strain gradient plasticity theories. J. Appl. Mech. {\bf 82} 071002-1-10.
\bibitem{FW1} Fleck, N.A.  and Willis, J.R. (2009). A mathematical basis for strain-gradient plasticity - Part I: Scalar plastic multiplier. J. Mech. Phys. Solids {\bf 57} 151--177.
\bibitem{FW2} Fleck, N.A.  and Willis, J.R. (2009). A mathematical basis for strain-gradient plasticity - Part II: Tensorial plastic multiplier. J. Mech. Phys. Solids {\bf 57} 1045-1057.
\bibitem{GHN} Gao H., Huang Y. and Nix W.D. 1999. Mechanism-based strain gradient plasticity ? I. Theory. J. Mech. Phys. Solids {\bf 47} 1239-1263.
\bibitem{GHNH} Gao H., Huang Y., Nix W.D. and Hutchinson J.W. (1999). Modeling plasticity at the micrometer scale, Naturwissenschaften {\bf 86} 507-515.
\bibitem{Gudmundson} Gudmundson, P. (2004). A unified treatment of strain gradient plasticity. J. Mech. Phys. Solids {\bf 52} 1379--1406.
\bibitem{GA} Gurtin, M.E., Anand, L. (2005). A theory of strain-gradient plasticity for isotropic, plastically irrotational materials. Part I: small
deformations. J. Mech. Phys. Solids {\bf 53} 1624--1649.
\bibitem{HR} Han, W. and Reddy, B.D. (2013). Plasticity: Mathematical Theory and Numerical Analysis. Springer, New York and Berlin.
\bibitem{ACE}
Korelc, J. (2002).  Multi-language and multi-environment generation of nonlinear finite element codes. Engng Comp. {\bf 18}(4) 312--327.
\bibitem{R} Reddy, B.D. (2011). The role of dissipation and defect energy in variational formulations of problems in strain-gradient plasticity. Part 2: single-crystal plasticity. Continuum Mech. Thermodyn. {\bf 23} 547--549.
\bibitem{REM} Reddy, B.D., Ebobisse, F. and McBride, A.T (2008). Well-posedness of a model of strain gradient plasticity for plastically irrotational materials. Int. J. Plast. {\bf 24} 55--73.
\bibitem{Rockafellar} Rockafellar, R.T. (1970). Convex Analysis. Princeton University Press, Princeton, N.J.
\bibitem{SKT} Simo, J.C., Kennedy J.G.  and Taylor, R.L. (1989). Complementary mixed finite element formulations for elastoplasticity. Comp. Meths  Appl. Mech. Engng {\bf 74}(2) 177--206.
\end{thebibliography}
\end{document}